\documentclass[aps,prd,twocolumn,superscriptaddress,floatfix]{revtex4-1}

\usepackage{textcomp,gensymb}
\usepackage{hyperref}
\usepackage{graphicx}
\usepackage{amsfonts,amsmath,amssymb,bm,bbm,latexsym}
\usepackage{color,soul}
\usepackage{slashed}
\usepackage[toc,page]{appendix}
\usepackage{flushend}
\usepackage{physics}
\usepackage[dvipsnames]{xcolor}
\usepackage{dcolumn}
\usepackage{float}
\usepackage{aas_macros}
\usepackage{fontawesome} 
\newcommand{\CO}{\mathcal{O}}

\newcommand{\githubmaster}{\href{https://github.com/neal-ak/supercaptngen}{\faGithub}}
\graphicspath{{./figures/}}

\hypersetup{
    pdfnewwindow=true,      
    colorlinks=true,       
    linkcolor=blue,          
    citecolor=blue,        
    filecolor=blue,      
    urlcolor=blue        
}

\begin{document}

\title{Dark Matter from Monogem}

\author{Christopher V. Cappiello}
\email{cvc1@queensu.ca}
\affiliation{Department of Physics, Engineering Physics, and Astronomy, Queen's University, Kingston, Ontario, K7N 3N6, Canada}
\affiliation{Arthur B. McDonald Canadian Astroparticle Physics Research Institute, Kingston ON K7L 3N6, Canada}
\affiliation{Perimeter Institute for Theoretical Physics, Waterloo ON N2L 2Y5, Canada}

\author{Neal~P. Avis~Kozar}
\email{neal.aviskozar@queensu.ca}
\affiliation{Department of Physics, Engineering Physics, and Astronomy, Queen's University, Kingston, Ontario, K7N 3N6, Canada}
\affiliation{Arthur B. McDonald Canadian Astroparticle Physics Research Institute, Kingston ON K7L 3N6, Canada}

\author{Aaron C. Vincent}
\email{aaron.vincent@queensu.ca}
\affiliation{Department of Physics, Engineering Physics, and Astronomy, Queen's University, Kingston, Ontario, K7N 3N6, Canada}
\affiliation{Arthur B. McDonald Canadian Astroparticle Physics Research Institute, Kingston ON K7L 3N6, Canada}
\affiliation{Perimeter Institute for Theoretical Physics, Waterloo ON N2L 2Y5, Canada}

\date{\today}

\begin{abstract}
As a supernova shock expands into space, it may collide with dark matter particles, scattering them up to velocities more than an order of magnitude larger than typical dark matter velocities in the Milky Way. If a supernova remnant is close enough to Earth, and the appropriate age, this flux of high-velocity dark matter could be detectable in direct detection experiments, particularly if the dark matter interacts via a velocity-dependent operator. This could make it easier to detect light dark matter that would otherwise have too little energy to be detected. We show that the Monogem Ring supernova remnant is both close enough and the correct age to produce such a flux, and thus we produce novel direct detection constraints and sensitivities for future experiments. \githubmaster
\end{abstract}

\maketitle


\section{Introduction}\label{sec:intro}
Identifying dark matter (DM) is one of the most important problems in particle physics and cosmology today, because although it makes up the vast majority of the matter in the universe, its particle properties remain a mystery \cite{Bertone:2004pz,Feng:2010gw,Buckley:2017ijx}. Decades of searching with direct detection experiments have produced strong constraints, but no unambiguous detection \cite{Cushman:2013zza,Schumann:2019eaa,Akerib:2022ort}. This lack of a discovery has motivated searches for DM beyond the traditional Weakly Interacting Massive Particle (WIMP) paradigm \cite{Knapen:2017xzo,Green:2020jor,Adams:2022pbo,Carney:2022gse,Mitridate:2022tnv}.

One class of DM model that could have avoided detection is DM that is strongly interacting, but lighter than $\sim$1 GeV. Such light particles, if traveling at typical halo velocities, carry too little momentum to trigger many of the most sensitive DM detectors. In recent years, this parameter space has been probed in several ways, including building lower threshold detectors, \cite{CRESST:2015txj,CRESST:2017ues,Collar:2018ydf,CRESST:2019jnq,EDELWEISS:2019vjv,SuperCDMS:2020aus,Mitridate:2022tnv,SuperCDMS:2022kse}, taking advantage of the Migdal effect or bremsstrahlung \cite{Kouvaris:2016afs,Ibe:2017yqa,Baxter:2019pnz,Bell:2019egg,EDELWEISS:2019vjv,Essig:2019xkx,XENON:2019zpr,GrillidiCortona:2020owp,Knapen:2020aky,2022arXiv220303993A}, and considering mechanisms that could produce higher-velocity DM. This last category includes upscattering by cosmic rays~\cite{Bringmann:2018cvk,Ema:2018bih,Bondarenko:2019vrb,Cappiello:2019qsw,Dent:2019krz,PROSPECT:2021awi,CDEX:2022fig,Maity:2022exk,Super-Kamiokande:2022ncz}, solar or supernova neutrinos~\cite{Zhang:2020nis,Das:2021lcr,Lin:2022dbl,Carenza:2022som}, blazars~\cite{Wang:2021jic,Granelli:2022ysi}, or energetic particles in the Sun~\cite{An:2017ojc,Emken:2017hnp}; annihilation or decay of heavy DM into lighter states~\cite{Agashe:2014yua,Berger:2014sqa,Kong:2014mia,Cherry:2015oca,Kopp:2015bfa,Alhazmi:2016qcs,Bhattacharya:2016tma,Necib:2016aez,Super-Kamiokande:2017dch}; primordial black hole evaporation~\cite{Marfatia:2022jiz}; acceleration of millicharged particles in supernova shocks~\cite{Hu:2016xas,Dunsky:2018mqs,Li:2020wyl}; and gravitational effects~\cite{Besla:2019xbx,Herrera:2021puj}.

We present a new mechanism for producing high-velocity DM: the upscattering of DM by nuclei in supernova shocks. Supernova shocks can expand at velocities in excess of 0.01$c$ (e.g. \cite{Sato:2016slx}), an order of magnitude larger than typical DM velocities at Earth's position in the Milky Way. If DM is light enough, collisions can boost it up to roughly double this expansion velocity, producing a small flux of light but high-momentum particles. We focus here on the specific case of the Monogem Ring supernova remnant. As shown below, the ratio of the Monogem Ring's distance from Earth to its age is comparable to the upscattered DM velocity, meaning such a flux would be visible today. Note that this mechanism is entirely distinct from Fermi acceleration, which, as mentioned above, could accelerate millicharged DM in supernova shocks \cite{Hu:2016xas,Dunsky:2018mqs,Li:2020wyl}. We do not require DM to be charged, only to have nonzero elastic scattering cross section with nuclei.

Given that we consider DM interacting with large, but still nonrelativistic, velocities, this mechanism is ideal for probing DM with velocity-dependent interactions with nuclei. Specifically, nonrelativistic effective field theory (NREFT) \cite{Fitzpatrick:2012ix,Anand:2013yka} provides numerous effective operators beyond the traditional spin-independent/spin-dependent dichotomy, many of which produce cross sections that scale strongly with velocity. In this work, we produce new constraints on several of these operators, constraining new parameter space at low mass and large coupling. Furthermore, we show that detectors being deployed in the coming year will be sensitive to supernova-upscattered DM over a wide range of parameter space for these operators.

This paper is organized as follows. In Sec.~\ref{sec:NREFT}, we introduce nonrelativistic effective field theory, and discuss the velocity scaling of the operators presented. In Sec.~\ref{sec:supernova_ejecta}, we describe supernova expansion into the surrounding interstellar medium (ISM). In Sec.~\ref{sec:upscattering}, we compute the flux of supernova-upscattered DM reaching Earth. In Sec.~\ref{sec:data_analysis}, we compare DM-induced event rates to experimental data. In Sec.~\ref{sec:limits}, we present our results. In Sec.~\ref{sec:conclusions}, we give our conclusions.

\section{Nonrelativistic Effective Field Theory}\label{sec:NREFT}
 Dark matter direct detection searches are typically performed in the context of  spin-independent (SI) or spin-dependent (SD) cross sections. These are velocity- and momentum-independent operators that arise naturally from nonrelativistic reductions of many generic UV-complete models (see, e.g., Ref.~\cite{Jungman:1995df}). However, there exist models in which momentum-dependent operators can dominate over the constant operators, motivating a more general treatment of possible interactions \cite{Chang:2009yt,Fan:2010gt,Kumar:2013iva}. The now-standard framework for the nonrelativistic effective field theory of DM interactions (NREFT, also called nonrelativistic effective operators) was laid out in Ref.~\cite{Fitzpatrick:2012ix}, in which the various possible responses are combinations of the following four Galilean-invariant quantities:
\begin{eqnarray}
 && i\vec{q}, \ \ \ \ \vec{v}_N^\perp \equiv \vec{v}_N + \frac{\vec{q}}{2\mu_N} , \ \ \ \ \vec{S}_\chi, \ \ \ \ \vec{S}_N,
\end{eqnarray}
where $\vec{S}_\chi$ and $\vec{S}_N$ are respectively the DM and nucleon spins, $\mu_N$ is their reduced mass, $\vec q$ is the exchanged momentum in a collision, and $\vec v_N$ is the DM-nucleon relative velocity. Ref.~\cite{Fitzpatrick:2012ix} tabulates 11 such operators, including the standard spin-independent response as $\CO_1$ and the spin-dependent response as $\CO_4$. A few additional operators that do not appear in the above reference can also be constructed, as detailed in Ref.~\cite{Anand:2013yka}, resulting in a final list of 15 linearly independent operators appropriate for DM of spin 0 or $\frac{1}{2}$. In Table~\ref{tab:operators}, we list the form of each of these operators, as well as the velocity scaling of the total elastic scattering cross section each operator produces. In the third column, $v$ is the DM-nucleus relative velocity. Additional operators can be constructed for higher-spin DM~\cite{Catena:2019hzw, Gondolo:2020wge}, but we restrict our work to the operators listed here.

\begin{table}
\begin{tabular}{ c c c  } 
 \hline
{Operator} & Form & $v$-scaling of $\sigma$ \\ [0.5ex] 
 \hline\hline
 $\CO_1$ & $1_{\chi}1_{N}$ & $v^0$ \\ 
 $\textcolor{Mulberry}{\CO_3}$ & $\textcolor{Mulberry}{i\vec{S}_N \cdot (\frac{\vec{q}}{m_N} \times \vec{v}_N^\perp)}$ & $\textcolor{Mulberry}{v^4}$ \\ 
 $\CO_4$ & $\vec{S}_{\chi} \cdot \vec{S}_N$ & $v^0$ \\ 
 $\CO_5$ & $i\vec{S}_{\chi} \cdot (\frac{\vec{q}}{m_N} \cross \vec{v}_N^\perp)$ & $v^4$ \\
 \textcolor{Mulberry}{$\CO_6$} & $\textcolor{Mulberry}{(\vec{S}_{\chi} \cdot \frac{\vec{q}}{m_N})(\vec{S}_{N} \cdot \frac{\vec{q}}{m_N})}$ & $\textcolor{Mulberry}{v^4}$ \\
 $\CO_7$ & $\vec{S}_N \cdot \vec{v}_N^\perp$ & $v^2$ \\
 $\CO_8$ & $\vec{S}_{\chi} \cdot \vec{v}_N^\perp$ & $v^2$ \\
 $\CO_9$ & $i\vec{S}_\chi \cdot (\vec{S}_N \times \frac{\vec{q}}{m_N})$ & $v^2$ \\ 
 $\CO_{10}$ & $i \vec{S}_N \cdot \frac{\vec{q}}{m_N}$ & $v^2$ \\
 $\CO_{11}$ & $i \vec{S}_\chi \cdot \frac{\vec{q}}{m_N}$ & $v^2$ \\
 $\CO_{12}$ & $\vec{S}_\chi \cdot (\vec{S}_N \times \vec{v}_N^\perp)$ & $v^2$ \\
 $\CO_{13}$ & $i (\vec{S}_\chi \cdot \vec{v}_N^\perp  ) (  \vec{S}_N \cdot \frac{\vec{q}}{m_N})$ & $v^4$ \\
 $\CO_{14}$ & $i (\vec{S}_\chi \cdot \frac{\vec{q}}{m_N}  ) (  \vec{S}_N \cdot \vec{v}_N^\perp)$ & $v^4$ \\
 $\textcolor{Mulberry}{\CO_{15}}$ & $\textcolor{Mulberry}{- ( \vec{S}_\chi \cdot \frac{\vec{q}}{m_N}) ((\vec{S}_N \times \vec{v}_N^\perp) \cdot \frac{\vec{q}}{m_N})}$ & $\textcolor{Mulberry}{v^6}$ \\
 \hline \hline
\end{tabular}
\caption{\label{tab:operators}List of effective dark matter-nucleon interaction operators, and the zero-momentum scaling of the corresponding cross section with the DM-nucleus relative velocity $v$. The operators we consider in this work are denoted in purple text.}
\end{table}
The DM-nucleon velocity can be separated into a sum of the DM-nucleus velocity, and the relative velocity of each nucleon with respect to the nucleus. The internal dynamics can then be projected onto nuclear basis states. As such, each of the operators listed here couples to a different combination of nuclear basis states. To turn the DM-nucleon interactions into DM-\textit{nucleus} matrix element, this change of basis must be performed, and evaluated for each individual isotope using a nuclear shell model. This allows the cross section to be factored into a \textit{DM response function}, dependent on the DM-nucleus relative velocity $v$, and a \textit{nuclear response function} which suppresses interactions for finite momentum exchanges. Nuclear response functions were computed and tabulated in the form $e^{-2y}$ times a polynomial in $y$, with $y \propto q^2$, by \cite{Fitzpatrick:2012ix} and \cite{Catena:2015uha} for a majority of elements that we will consider here. 

From Table~\ref{tab:operators}, we see that almost all of the effective operators listed produce strongly velocity-dependent cross sections, with $\CO_{15}$ scaling as strongly as $v^6$. It is valuable to probe velocity-dependent operators at velocities significantly higher than the galactic escape velocity, as the interaction strength may be orders of magnitude larger than it would be at halo velocities. In our analysis, we focus on $\CO_{15}$, and on two operators, $\CO_{3}$ and $\CO_{6}$, that scale as $v^4$. 

\section{Supernova Ejecta}\label{sec:supernova_ejecta}
\subsection{Sedov-Taylor Solution}\label{ss:sedovtaylor}
The early stages of the expansion of a supernova shock into a uniform medium are described by the Sedov-Taylor blastwave solution \cite{1946JApMM..10..241S,1950RSPSA.201..159T,1950RSPSA.201..175T,1959sdmm.book.....S}. In the free expansion or ejecta-dominated phase, the mass of the swept up ISM is small compared to the ejecta mass, and has little effect on the expansion of the shock. The Sedov-Taylor phase begins when enough ejecta has been swept up to significantly impact the expansion of the shock, usually around 1000 years after the explosion. This causes the expansion to slow down more rapidly. 

The Sedov-Taylor solution allows us to compute the radius and expansion velocity of the shock as a function of time, for a given explosion energy, ejecta mass, and ambient ISM density. We adopt the parametrization given in Refs.~\cite{Cardillo:2015zda,Cristofari:2021hbc}, convenient for the discussion of supernova shocks: the radius is given by

\begin{eqnarray}\label{eq:rshock}
    R_s(t) = R_0\left(\left(\frac{t}{t_0}\right)^{-5\lambda_{FE}} + \left(\frac{t}{t_0}\right)^{-5\lambda_{ST}}\right)^{-1/5}\,,
\end{eqnarray}
and the shock velocity by
\begin{multline}\label{eq:vshock}
    V_s(t) = \frac{R_0}{t_0}\left(\frac{R_s(t)}{R_0}\right)^6 \times \\ \left(\lambda_{FE}\left(\frac{t}{t_0}\right)^{-5\lambda_{FE}-1} + \lambda_{ST}\left(\frac{t}{t_0}\right)^{-5\lambda_{ST}-1}\right)\,.
\end{multline}
Here $\lambda_{FE}$ and $\lambda_{ST}$ are parameters describing the free expansion and Sedov-Taylor phases, respectively.

 For a Type Ia supernova expanding into a uniform medium, $\lambda_{ST} = 2/5$, and $\lambda_{FE} = 4/7$. $R_0 = \left(\frac{3M_{ej}}{4\pi n_0(1.27m_p)}\right)^{1/3}$ is a scale radius, and $t_0 = \left(R_0\left(\frac{M_{ej}n_0(1.27m_p)}{0.38E_{SN}^2}\right)^{1/7}\right)^{7/4}$, with $M_{ej}$ denoting the ejecta mass, $E_{SN}$ the explosion energy, and $n_0$ the ambient ISM density. The factor of $1.27m_p$ is the average ISM mass per nucleus.

For a Type II supernova, expanding into the progenitor star's presupernova wind, $\lambda_{ST} = 2/3$, and $\lambda_{FE} = 6/7$. The scale radius is $R_0 = \frac{M_{ej}V_{w}}{\dot M}$, where $V_w$ is the wind velocity and $\dot M$ is the mass loss rate. Following \cite{Cardillo:2015zda}, we assume $V_w = 10$ km/s and $\dot M = 10^{-5} M_{\odot}$/year. In this case $t_0 = \left(R_0\left(\frac{\dot M}{36\pi}\frac{(18M_{ej})^{-5/2}}{(40E_{SN})^{-3/2}}(\frac{40E_{SN}}{18M_{ej}})^{-9/2}\right)^{1/7}\right)^{7/3}$. The density of the presupernova wind around the supernova remnant, which we will use in the following Section, is

\begin{equation}\label{eq:winddensity}
    \rho(r) = \frac{\dot M}{4\pi V_w r^2}\,.
\end{equation}

\subsection{Ejecta Composition}
We are also interested in the elemental composition of the ejecta, as different nuclei can in principle have very different scattering cross sections with DM. Ref.~\cite{Rauscher:2001dw} performed simulations to model nucleosynthesis in stars ranging from 15 to 25 $M_{\odot}$, and reported the abundance of different nuclear isotopes in the ejecta relative to solar abundances. In Table~\ref{tab:mass fractions}, we report the mass fractions of the most abundant nuclei in supernova ejecta, averaged from the results of five simulations reported by Ref.~\cite{Rauscher:2001dw} at different progenitor masses. The mass fractions are dominated by hydrogen and helium, as in the Sun, but the abundances of many heavier elements are enhanced by around a factor of 10 compared to solar abundances.

\begin{table}
\begin{tabular}{ l l} 
 \hline
Nucleus & $f_i$ \\ [0.5ex] 
 \hline\hline
$^1$H & 0.493 \\ 
$^4$He & 0.35 \\
 $^{16}$O & 0.1 \\
 $^{28}$Si & 0.02 \\
 $^{12}$C & 0.015 \\
 $^{56}$Fe & 0.007 \\
 $^{20}$Ne & 0.005 \\
 $^{24}$Mg & 0.005 \\
 $^{32}$S & 0.005 \\
 $^{14}$N & 0.004 \\
 $^{23}$Na & 0.0004 \\ [1ex] 
 \hline \hline
\end{tabular}
\caption{\label{tab:mass fractions}Approximate mass fractions for the most abundant nuclei in supernova ejecta, averaged from results of Ref.~\cite{Rauscher:2001dw} for progenitor masses from 15 to 25 $M_{\odot}$}
\end{table}

\subsection{The Monogem Ring}
The Monogem Ring is one of the closest supernova remnants to Earth, with an apparent diameter of 25$\degree$ on the sky. Its distance from Earth is believed to be approximately $D = 300$ parsecs, based on supernova energetics, as well as the parallax distance to both the (arguably) associated pulsar PSR B0656+14 and a seemingly interacting star cluster in the Gemini H $\alpha$ ring \cite{1996ApJ...463..224P,Thorsett:2003xy,2018MNRAS.477.4414K}. The most recent Sedov-Taylor calculation gives the Monogem Ring an age of 68,000 years, an explosion energy of $8.38\times10^{50}$ erg, and a surrounding ISM density of $3.73\times10^{-3}$ cm$^{-3}$~\cite{2018MNRAS.477.4414K}. These fits do not, however, provide a value for the ejecta mass. Inserting these fit values into Eqns.~\ref{eq:rshock} and ~\ref{eq:vshock} gives a present day radius and velocity that agree with the reported values to within a factor of 2, the values being nearly independent of the assumed ejecta mass.

\section{Dark Matter Upscattering in Supernova Shocks}\label{sec:upscattering}

\subsection{Dark Matter Velocity Distribution}

We assume that all of the ejecta produced by the supernova is located in a thin shell at $R_s(t)$ traveling at the shock velocity $V_s(t)$. While the density distribution within a supernova remnant is model dependent, it peaks sharply near the shock in the widely used Sedov model~\cite{1959sdmm.book.....S}, and even more sharply in the more recent Chevalier model~\cite{1982ApJ...258..790C}, with the result that the majority of the mass is indeed concentrated within $\sim$10\% of the shock radius. The number of DM particles that this shell encounters per unit time is 

\begin{equation}
    4\pi R_s(t)^2\frac{\rho_{\chi}}{m_{\chi}}V_s(t)\,,
\end{equation}
where we adopt the standard value $\rho_{\chi} = 0.3$ GeV cm$^{-3}$ (the distance to Monogem is small compared to the scale radius of the Milky Way halo). In the limit of an optically thin shell, meaning that the probability of a dark matter particle colliding with a nucleus is $\ll 1$, the probability that a dark matter particle is struck by a nucleus as the shell passes by it is given by

\begin{equation}
    \sum_i \frac{M_{ej}f_i}{m_i}\frac{1}{4\pi R_s(t)^2}\sigma_{\chi i}\,,
\end{equation}
where $m_i$ is the mass of a nucleus of species $i$, and the sum is over the nuclei in the ejecta.

Strictly speaking, the above formula only accounts for dark matter being struck by the ejecta. If we add to the shell the mass of the presupernova wind that it sweeps up as it expands, treating this ambient material as composed purely of hydrogen, the probability becomes

\begin{equation}
    \sum_i \left(\frac{M_{ej}f_i}{m_i} + 4\pi\int_0^{R_s(t)} n(r)r^2\mathrm{d}r\delta_{i,1}\right)\frac{1}{4\pi R_s(t)^2}\sigma_{\chi i}\,.
\end{equation}
where $n(r)$ is the density of the presupernova wind obtained from Eq.~\ref{eq:winddensity}, and $i=1$ denotes hydrogen, so that the delta function picks out only the hydrogen contribution.

To turn this into a differential rate, we  replace $\sigma_{\chi i}$ with $\frac{\mathrm{d}\sigma_{\chi i}}{dv}$, where we use $dE = m_{\chi}vdv$ to write

\begin{equation}
    \frac{\mathrm{d}\sigma_{\chi i}}{\mathrm{d}v} = m_{\chi}v\frac{\mathrm{d}\sigma_{\chi i}}{\mathrm{d}E}\,.
\end{equation}
This allows us to write

\begin{multline}
    \Phi(v,t) = \int \mathrm{d}E\, \delta(E-\frac{1}{2}m_{\chi}v^2)\rho_{\chi}V_s(t)\times \\
    \sum_i \left(\frac{M_{ej}f_i}{m_i} + 4\pi\int_0^{R_s(t)} n(r)r^2\mathrm{d}r\delta_{i,1}\right)v\frac{\mathrm{d}\sigma_{\chi i}}{\mathrm{d}E}\,,
\end{multline}
where the factor of $m_{\chi}$ has canceled out, though the differential cross section still implicitly depends on the mass. To avoid confusion, we note that the factor of $V_s$ here represents the rate at which the ejecta is passing through the DM, while the factor of $v$ arises from the transformation of the differential cross section above.

The flux at Earth can then be obtained by integrating this flux over all time, ensuring that at a given time $t$, only particles upscattered to a velocity $v = D/(Age-t)$ contribute: particles at any other velocity would have either already passed the Earth, or not reached us yet. Accounting for the geometric factor of $4\pi D^2$, this yields,

\begin{figure}[t]
    \centering
    \includegraphics[width=\columnwidth]{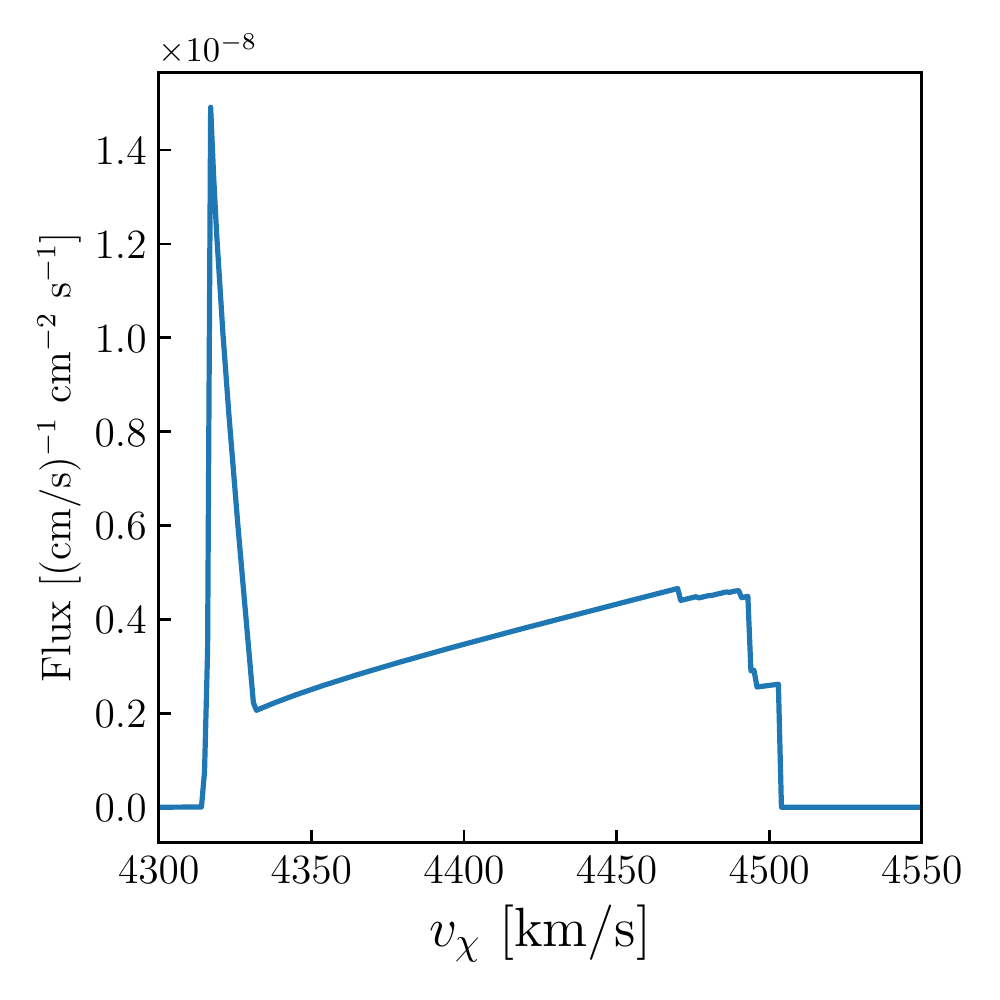}
    \caption{Flux of upscattered DM particles reaching Earth as a function of the DM velocity $v$, for operator $\CO_{15}$, at a coupling of $c_{15}^2m_v^4 = 7\times 10^{28}$ and $m_{\chi} = 1$ GeV. Here $m_v = $ 246 GeV is the Higgs vev.}
    \label{fig:fluxatearth}
\end{figure}

\begin{equation}\label{eq:earthflux}
    \Phi_{Earth}(v) = \frac{1}{4\pi D^2}\int \Phi(v,t) \delta(t - (Age - D/v)) \mathrm{d}t\,.
\end{equation}

\subsection{Numerical Implementation}

We compute the DM flux at Earth (Eq.~\ref{eq:earthflux}) in a general NREFT framework using a modified version of the publicly available Capt'n General code~\cite{GAMBIT:2021rlp,Kozar:2021iur}. This code was originally developed to model DM capture in the Sun in effective theories. In this work, we have modified it to model DM being upscattered by stellar material to a velocity $v$, based on the equations in the preceding Sections, rather than being downscattered to below stellar escape velocity. Interactions could in principle be described by an arbitrary combination of the operators listed in Table~\ref{tab:operators}, for both isoscalar and isovector interactions, though we consider only individual operators and only the isoscalar case for this work.

For the case of spin-independent scattering ($\CO_1$), we validate our results by independently computing the flux at both Monogem and Earth, and find no significant difference between the two implementations.

Fig.~\ref{fig:fluxatearth} shows an example DM flux at Earth for $m_{\chi} = 1$ GeV. The different steps are due to scattering with different nuclei, which leads to a maximum DM velocity that depends on the nucleus mass. The large peak at low velocity comes from scattering with hydrogen, while the subsequent small peaks are due to helium, carbon, nitrogen, etc. The general trend of rising flux with higher velocity is due to the differential cross section's strong dependence on momentum transfer, while the downward slope after each step is due to the range of supernova ages (and thus ejecta velocities) that contribute to the upscattered flux. Note that the sharpness of the peaks is due to our assumption that all the ejecta travels at the same velocity, and is not an artifact of the plot's velocity resolution. When setting limits, the sharpness of these peaks is blunted by the detector energy resolution and the range of allowed recoil energies, so such a sharply peaked velocity distribution should not lead to unphysically narrow recoil spectra. 

\section{Data Analysis}\label{sec:data_analysis}

\subsection{Direct Detection Rates}

Given a DM flux, mass, and differential cross section, it is possible to compute the recoil spectrum in a detector. The DM flux $\Phi_{Earth}(v)$, with units of (cm$^{-2}$ s$^{-1}$)(cm/s)$^{-1}$, plays the role that the quantity $\frac{\rho_{\chi}}{m_{\chi}} v f(v)$ would play in traditional direct detection formalism. In other words, the detection rate can be given by

\begin{equation}\label{eq:recoilspectrum}
    \frac{\mathrm{d}R}{\mathrm{d}E_r} = \frac{1}{m_N}\int_{v_{min}(E_r)}^{\infty}\mathrm{d}v\, \Phi_{Earth}(v)\frac{\mathrm{d}\sigma_{\chi N}}{\mathrm{d}E_r}(v)\,,
\end{equation}
where $N$ is the nucleus making up the detection material.

To compute the resulting recoil spectrum in the NREFT framework, we use a slightly modified version of the publicly available \texttt{WIMpy\_NREFT} code~\cite{WIMpy-code}. This code computes recoil spectra for numerous nuclei and for an arbitrary combination of effective operators, and has been used by experimental collaborations such as DEAP~\cite{DEAP:2020iwi}.

\subsection{Detectors Used}

Because $\Phi_{Earth}$ is proportional to the differential cross section, the recoil spectrum in Eq.~\ref{eq:recoilspectrum} scales with two factors of the cross section. This means that increasing the cross section rapidly increases the detectability of the supernova-upscattered DM. For this reason, the best existing detectors to search for supernova upscattered DM are surface detectors, which have little shielding other than the atmosphere. In this section, we use data from the athermal phonon detector operated by the SuperCDMS Collaboration~\cite{SuperCDMS:2020aus}, because of its low energy threshold, location on the Earth's surface, and its relatively large exposure compared to other surface detectors such as the CRESST surface run~\cite{CRESST:2017ues}. (This SuperCDMS detector does have some copper shielding, which will be discussed below.) 

\begin{figure}[t]
    \centering
    \includegraphics[width=\columnwidth]{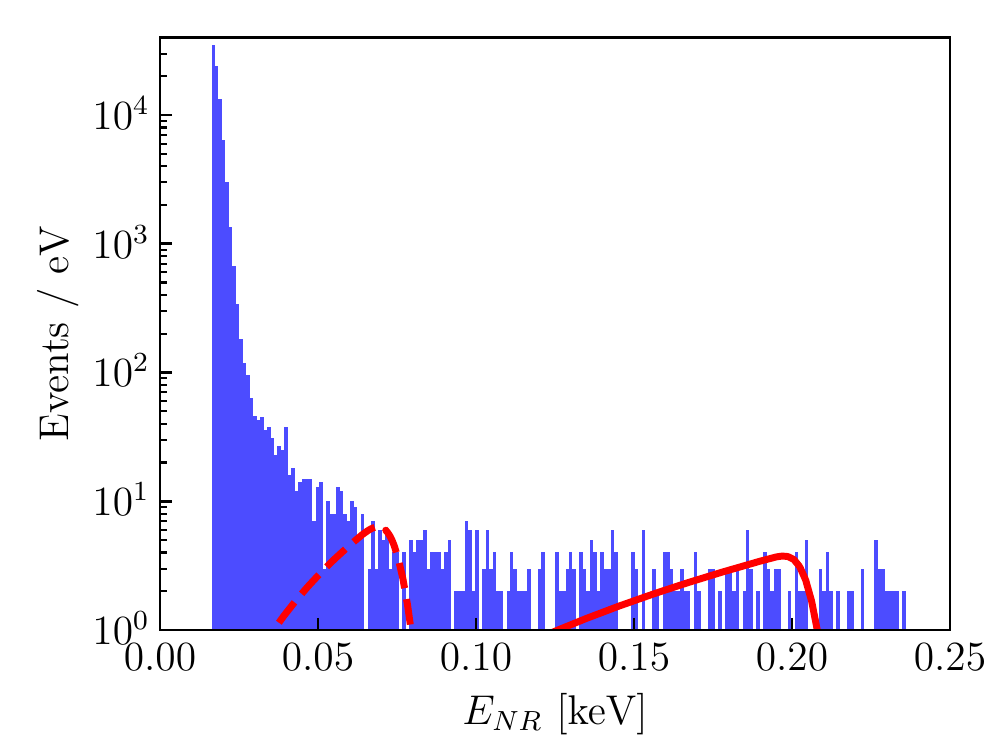}
    \caption{Data from the SuperCDMS surface run~\cite{SuperCDMS:2020aus} (blue histogram), compared to the event spectra for DM with $m_{\chi} =$ 100 MeV (dashed red) and $m_{\chi} =$ 1 GeV (solid red). Both DM spectra are for operator $\CO_{15}$, at couplings that are just barely ruled out by the data, $c_{15}^2m_v^4 = 5\times 10^{32}$ and $c_{15}^2m_v^4 = 7\times 10^{28}$, respectively.}
    \label{fig:cdmsdata}
\end{figure}

However, as the surface detector target material is silicon, it has no sensitivity to purely spin-dependent operators such as $\CO_6$. For this reason, we also derive limits from PICO-60~\cite{PICO:2019vsc}, which is particularly sensitive to spin-dependent interactions. Although PICO-60 is deep underground, the nuclei that make up most of Earth's crust do not interact via $\CO_6$, so the shielding is primarily due to nitrogen in the atmosphere, and to subdominant components of Earth's crust like aluminum and sodium. 

We furthermore find that a future underground detector with low threshold, low background, and large exposure would be able to probe a large region of DM parameter space. Specifically, we focus on the SuperCDMS germanium HV detectors, planned for deployment at SNOLAB. The projected exposure for these detectors is 36 kg-yr~\cite{SuperCDMS:2022kse}, compared to the gram-day scale exposure of the existing surface run~\cite{SuperCDMS:2020aus}. For the Ge HV detectors, we assume a detection threshold of 40 eVnr as reported in Ref.~\cite{SuperCDMS:2016wui}, and take the projected background from this same reference, with a rate of 1--10 events/kg/yr/keVnr for most of the ROI.

Finally, we can also compute projected limits for a large, future xenon experiment like DARWIN. If no candidate events are seen in a 200 ton-year exposure~\cite{DARWIN:2016hyl}, and assuming the same energy range and efficiency as XENON1T~\cite{XENON:2018voc}, such a detector could also set limits on supernova-boosted DM.

\subsection{Dark Matter Propagation to the Detector}

The couplings that we aim to probe are large enough that DM particles would scatter in the Earth's crust or atmosphere before reaching a detector, deflecting them away from their original trajectory and causing them to lose energy. A commonly used formalism to model this attenuation is to assume that DM travels in a straight line to the detector, modeling its energy loss as a continuous process, with the energy-loss rate

\begin{equation}
    \frac{\mathrm{d}E}{\mathrm{d}x} = -\sum_i n_i \int_0^{E_{max,i}}\frac{\mathrm{d}\sigma_{\chi i}}{\mathrm{d}E}\,E \,\mathrm{d}E\,,
\end{equation}
where the summation is over nuclei in the crust or atmosphere. 

This formalism is not strictly correct when modeling light DM, as particles may scatter through large angles, making the straight-line assumption inaccurate. However, Ref.~\cite{Emken:2018run} showed that this straight-line approximation produces reasonably accurate or even conservative results when compared to more complete numerical simulations. As a result, the publicly available \texttt{VERNE} code~\cite{verne,Kavanagh:2017cru} has been used for multiple low-mass direct detection analyses, where the scattering is essentially isotropic~\cite{EDELWEISS:2019vjv,XENON:2019zpr,SuperCDMS:2020aus}. In fact, because we consider cross sections that scale strongly with velocity, assuming that all particles lose the same amount of energy is conservative, as particles that lose less than the average amount of energy will interact with a larger cross section once reaching the detector.

\begin{figure}[!ht]
\hspace*{-1cm}\includegraphics[width=1.0\columnwidth]{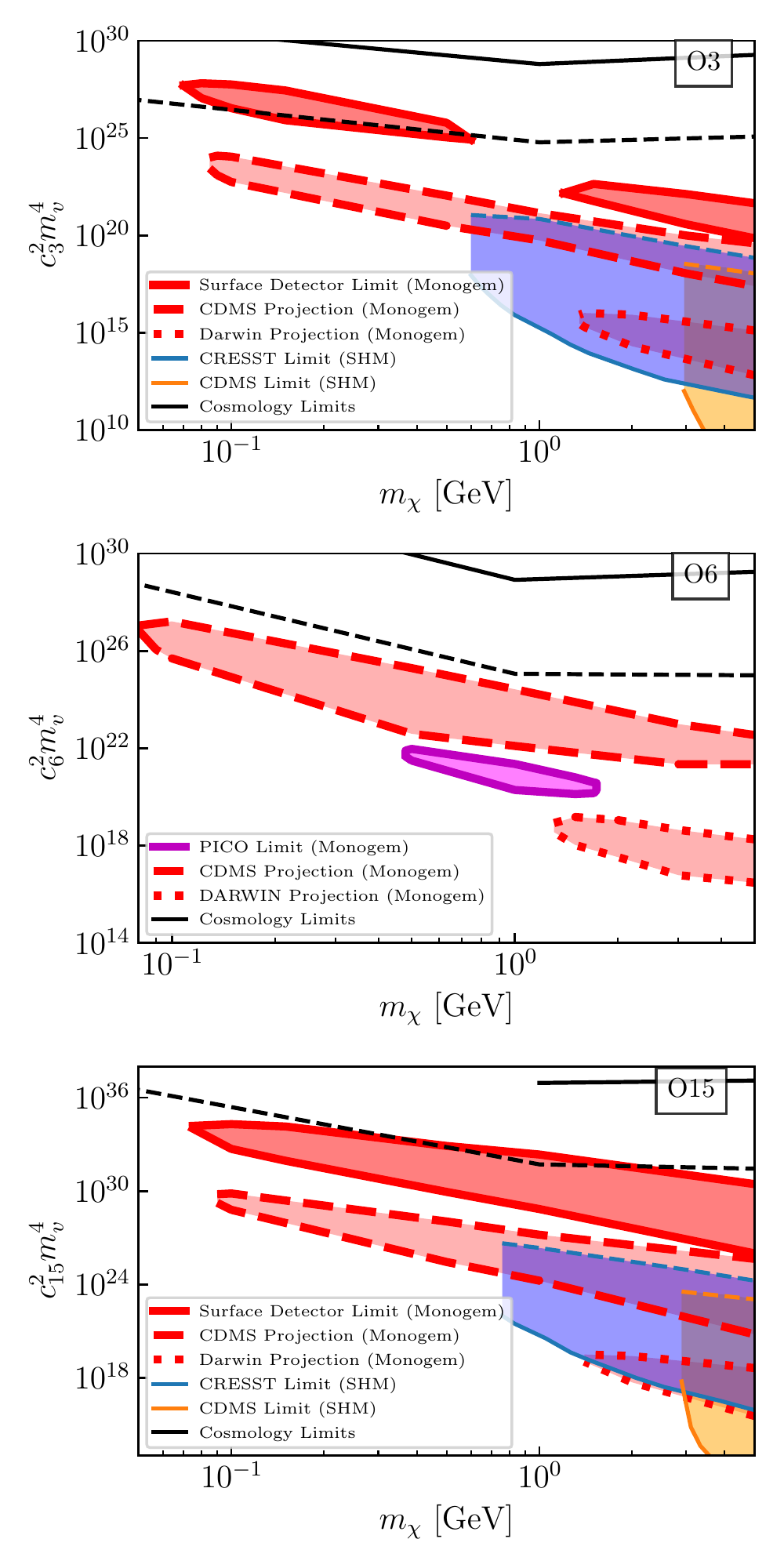}
    \caption{Limits on three non-relativistic operators due to the high-velocity dark matter component from the Monogem Supernova remnant. Results of this paper are shown in red and purple, where the lower edges are limited by the experiment sensitivity, and the upper edges by the overburden attenuation of DM flux on its way to the detector. Standard Halo Model (SHM) limits from CRESST-II (blue) are taken from Ref.~\cite{CRESST:2018vwt}, and limits from CDMS (orange) are from Ref.~\cite{SuperCDMS:2015lcz}. The black cosmological limits are based on Milky Way satellites~\cite{Maamari:2020aqz} (dashed) and CMB damping~\cite{Boddy:2018kfv} (solid).}
    \label{fig:limits}
\end{figure}

We assume that any flux arriving from below the horizon is completely stopped by the Earth, and that Monogem is below the horizon for exactly half the day (it is actually below the horizon for very slightly less than half the day, so this is mildly conservative). We further assume that all particles arriving from above the horizon arrive at an angle 45$\degree$ from vertical, as this is roughly Monogem's median zenith angle at the latitudes of the detectors we consider. This is again conservative, as the particles that traverse less shielding would arrive with much larger velocities and thus cross sections at the detector.

For deep-underground detectors like PICO-60, the SuperCDMS SNOLAB experiment, and DARWIN, we neglect atmospheric shielding, considering only elements in the Earth's crust. We take the crust composition from Ref.~\cite{Emken:2018run}. For the SuperCDMS surface run, we consider $\sim$14 meters water equivalent (mwe) of atmospheric shielding (i.e. the 10 mwe thickness of the atmosphere divided by $\cos 45 \degree$). The surface run also has 5 cm of copper around it, but unfortunately we have not been able to find differential cross sections for copper tabulated for the effective operators in question. Instead, we have computed cross sections for several different isotopes of iron, nickel, germanium, silicon, and aluminum, and opted to conservatively model copper as $^{56}$Fe, as this nucleus had the largest cross section of any of the nuclei we considered.

For a more complete discussion of attenuation in the analogous case of cosmic ray-upscattered dark matter, see Refs.~\cite{Bringmann:2018cvk,Cappiello:2019qsw,Alvey:2022pad}.

\subsection{Limit Setting}

Following Ref.~\cite{SuperCDMS:2020aus}, we set limits from the SuperCDMS surface run using the optimal interval method~\cite{Yellin:2002xd}. This method assumes that all observed events could in principle be dark matter, so no background subtraction or fitting is performed. Roughly speaking, a dark matter mass-cross section pair is ruled out when the corresponding recoil spectrum is significantly larger than the observed event rate over a properly optimized energy interval. 

Fig.~\ref{fig:cdmsdata} shows the surface run data compared to the predicted DM spectrum for a $m_{\chi} = 1$ GeV and $c_{15}^2m_v^4 = 7\times 10^{28}$, a coupling that is just barely ruled out by the data. As with the velocity distribution, the recoil spectrum rises with energy because of the dependence on momentum transfer. Increasing the coupling increases the height of the recoil spectrum, but also pushes it to lower energy due to attenuation. If the coupling is made too large, the recoil spectrum will be either pushed entirely below the detector's energy threshold, or hidden by the high background at low energy.

Ref.~\cite{PICO:2019vsc} reports a total of 3 DM candidate events in the full PICO-60 data set. We determine the 90\% CL by computing the range of mass and cross section that would produce an expectation value of 6.7 events in the full PICO-60 exposure.

The projection for DARWIN is determined in a similar way: assuming a 200 ton-year exposure~\cite{DARWIN:2016hyl}, and the same efficiency and energy range as XENON1T~\cite{XENON:2018voc}, we compute the region of parameter space with an expectation value of at least 2.3 events, the 90\% CL if zero candidate events are observed.

Finally, for the SuperCDMS SNOLAB projections, we perform a profile likelihood analysis of artificial data generated from projected backgrounds. For each DM mass and coupling, we generate 20 artificial data sets based on the projected background reported in Ref.~\cite{SuperCDMS:2016wui}. For each data set we perform a profile likelihood analysis, in the form of an unconstrained fit over the six projected background spectra plus the signal spectrum. We rule out a mass-coupling pair if at least 90\% of the data sets rule it out at the 90\% level.

\section{Results}\label{sec:limits}

Fig.~\ref{fig:limits} shows our limits and projections for three effective operators: $\CO_{3}$, $\CO_{6}$, and $\CO_{15}$. The total cross sections for $\CO_{3}$ and $\CO_{6}$ scale with $v^4$, with $\CO_{6}$ being representative of spin-dependent interactions, while the cross section for $\CO_{15}$ scales with $v^6$. 

Our limits from the SuperCDMS surface run reach down to masses of about 70 MeV, about 20 MeV lower than limits based on the virialized Standard Halo Model (SHM) dark matter using the same detector. The limit from PICO-60 extends to just below 0.5 GeV, compared to the minimum mass of $\sim$3 GeV excluded by the SHM PICO-60 analysis. The large gap in our surface run exclusion region for $\CO_{3}$, from about 0.6--1.2 GeV is due to the shape of the observed data. The slope of the observed recoil spectrum is such that the signal spectrum for this mass range is always hidden by the data. Lighter particles lose energy very inefficiently during attenuation, while heavier particles carry more energy to begin with, meaning both lighter and heavier particles can be excluded for some range of couplings.

The projections show similar behavior. The projections for SuperCDMS extend down to 80--90 MeV, compared to a minimum mass of 400--500 MeV for the SHM analysis. The projections for DARWIN reach a minimum mass of 1.35--1.4 GeV, down from $\sim$5 GeV for a SHM analysis.

Several direct detection experiments have set existing limits on effective operators, including CDEX~\cite{CDEX:2020tkb}, CRESST-II~\cite{CRESST:2018vwt}, DarkSide-50~\cite{DarkSide-50:2020swd}, DEAP-3600~\cite{DEAP:2020iwi}, LUX~\cite{LUX:2020oan,LUX:2021ksq}, PandaX-II~\cite{PandaX-II:2018woa}, SuperCDMS Soudan and CDMSLite~\cite{SuperCDMS:2015lcz,SuperCDMS:2022crd}, and XENON100~\cite{XENON:2017fdd}. However, most of these limits do not extend below a few GeV, and most do not constrain all operators. For $\CO_{3}$ and $\CO_{15}$, we show limits from CRESST-II~\cite{CRESST:2018vwt} and CDMS~\cite{SuperCDMS:2015lcz}, along with estimated ceilings for both experiments. 

Finally, cosmological and astrophysical limits have been set on effective operators or, more generally, velocity-dependent scattering. Ref.~\cite{Boddy:2018kfv} set limits on effective operators based on DM scattering with hydrogen and helium leading to damping in the CMB at small scales. Ref.~\cite{Maamari:2020aqz} set limits on general velocity-dependent scattering based on the Milky Way satellite galaxy population, which would be suppressed in the presence of such DM-nucleus interactions. Though these limits are not explicitly limits on effective field theory parameters, we recast them as limits on the corresponding operators. We show constraints from both these references in Fig.~\ref{fig:limits}.

In principle, more detailed modeling of the ejecta's velocity structure could impact our limits. However, we expect this to be a small effect, especially at the low masses we are most interested in. As a cross check, we have computed how our limits would change if Monogem's expansion were treated as that of a Type Ia supernova rather than Type II, modifying the equations of Sec.~\ref{ss:sedovtaylor} according to Ref.~\cite{Cardillo:2015zda,Cristofari:2021hbc}. Qualitatively, the largest change is that the shock is expanding into the ambient ISM rather than a dense presupernova wind, resulting in higher velocity but less swept up hydrogen (in both cases we assume 5 M$_{\odot}$ of ejecta, admittedly too large for a Type Ia supernova, but our interest here was just on the impact of the velocity evolution). The limits we obtained in that case were slightly stronger for $m_{\chi} \gtrsim 1$ GeV, but virtually unchanged for lower masses. Because our results are not strongly affected by such a large change, we expect that the effect of a wider distribution of ejecta velocities would likewise be small. Further, a more realistic distribution of ejecta velocities would not necessarily lead to a less sharply peaked DM velocity distribution, because the shock, by its very nature, produces a sharp jump in both density and velocity.

\section{Conclusions}\label{sec:conclusions}
Supernovae are among the most energetic events in the Universe. In this work, we considered the Monogem Ring supernova remnant as a source of upscattered dark matter, as has been previously done for cosmic rays, the Sun, blazars, and supernova neutrinos. Exploiting the high---but still nonrelativistic---velocity of such dark matter, we derive limits and projected sensitivities to light dark matter that scatters via velocity-dependent effective operators. Our results extend to lower mass than existing nonrelativistic effective field theory searches, and cover a wide range of couplings. Our work provides a new avenue for future experiments to search for light dark matter that would otherwise be doomed to lurk below threshold.

\acknowledgements{}
We are grateful to Katie Auchettl, John Beacom, Riccardo Catena, Pierre Cristofari, and Todd Thompson for helpful discussions, and especially to Bradley Kavanagh for updates made to, and help with, the \texttt{WIMpy\_NREFT} code. We also thank the anonymous referee for their helpful suggestions, which improved the paper.
The authors are supported by the Arthur B. McDonald Canadian Astroparticle Physics Research Institute, NSERC, and the Province of Ontario through an Early Research Award. Computations were performed on equipment funded by the Canada Foundation for Innovation and the Ontario Government and operated by the Queen's Centre for Advanced Computing.
Research at Perimeter Institute is supported by the Government of Canada through the Department of Innovation, Science, and Economic Development, and by the Province of Ontario.

\bibliography{supernova}

\begin{thebibliography}{104}%
\makeatletter
\providecommand \@ifxundefined [1]{%
 \@ifx{#1\undefined}
}%
\providecommand \@ifnum [1]{%
 \ifnum #1\expandafter \@firstoftwo
 \else \expandafter \@secondoftwo
 \fi
}%
\providecommand \@ifx [1]{%
 \ifx #1\expandafter \@firstoftwo
 \else \expandafter \@secondoftwo
 \fi
}%
\providecommand \natexlab [1]{#1}%
\providecommand \enquote  [1]{``#1''}%
\providecommand \bibnamefont  [1]{#1}%
\providecommand \bibfnamefont [1]{#1}%
\providecommand \citenamefont [1]{#1}%
\providecommand \href@noop [0]{\@secondoftwo}%
\providecommand \href [0]{\begingroup \@sanitize@url \@href}%
\providecommand \@href[1]{\@@startlink{#1}\@@href}%
\providecommand \@@href[1]{\endgroup#1\@@endlink}%
\providecommand \@sanitize@url [0]{\catcode `\\12\catcode `\$12\catcode
  `\&12\catcode `\#12\catcode `\^12\catcode `\_12\catcode `\%12\relax}%
\providecommand \@@startlink[1]{}%
\providecommand \@@endlink[0]{}%
\providecommand \url  [0]{\begingroup\@sanitize@url \@url }%
\providecommand \@url [1]{\endgroup\@href {#1}{\urlprefix }}%
\providecommand \urlprefix  [0]{URL }%
\providecommand \Eprint [0]{\href }%
\providecommand \doibase [0]{http://dx.doi.org/}%
\providecommand \selectlanguage [0]{\@gobble}%
\providecommand \bibinfo  [0]{\@secondoftwo}%
\providecommand \bibfield  [0]{\@secondoftwo}%
\providecommand \translation [1]{[#1]}%
\providecommand \BibitemOpen [0]{}%
\providecommand \bibitemStop [0]{}%
\providecommand \bibitemNoStop [0]{.\EOS\space}%
\providecommand \EOS [0]{\spacefactor3000\relax}%
\providecommand \BibitemShut  [1]{\csname bibitem#1\endcsname}%
\let\auto@bib@innerbib\@empty
\bibitem [{\citenamefont {Bertone}\ \emph {et~al.}(2005)\citenamefont
  {Bertone}, \citenamefont {Hooper},\ and\ \citenamefont
  {Silk}}]{Bertone:2004pz}%
  \BibitemOpen
  \bibfield  {author} {\bibinfo {author} {\bibfnamefont {G.}~\bibnamefont
  {Bertone}}, \bibinfo {author} {\bibfnamefont {D.}~\bibnamefont {Hooper}}, \
  and\ \bibinfo {author} {\bibfnamefont {J.}~\bibnamefont {Silk}},\ }\href
  {\doibase 10.1016/j.physrep.2004.08.031} {\bibfield  {journal} {\bibinfo
  {journal} {Phys. Rept.}\ }\textbf {\bibinfo {volume} {405}},\ \bibinfo
  {pages} {279} (\bibinfo {year} {2005})},\ \Eprint
  {http://arxiv.org/abs/hep-ph/0404175} {arXiv:hep-ph/0404175} \BibitemShut
  {NoStop}%
\bibitem [{\citenamefont {Feng}(2010)}]{Feng:2010gw}%
  \BibitemOpen
  \bibfield  {author} {\bibinfo {author} {\bibfnamefont {J.~L.}\ \bibnamefont
  {Feng}},\ }\href {\doibase 10.1146/annurev-astro-082708-101659} {\bibfield
  {journal} {\bibinfo  {journal} {Ann. Rev. Astron. Astrophys.}\ }\textbf
  {\bibinfo {volume} {48}},\ \bibinfo {pages} {495} (\bibinfo {year} {2010})},\
  \Eprint {http://arxiv.org/abs/1003.0904} {arXiv:1003.0904 [astro-ph.CO]}
  \BibitemShut {NoStop}%
\bibitem [{\citenamefont {Buckley}\ and\ \citenamefont
  {Peter}(2018)}]{Buckley:2017ijx}%
  \BibitemOpen
  \bibfield  {author} {\bibinfo {author} {\bibfnamefont {M.~R.}\ \bibnamefont
  {Buckley}}\ and\ \bibinfo {author} {\bibfnamefont {A.~H.~G.}\ \bibnamefont
  {Peter}},\ }\href {\doibase 10.1016/j.physrep.2018.07.003} {\bibfield
  {journal} {\bibinfo  {journal} {Phys. Rept.}\ }\textbf {\bibinfo {volume}
  {761}},\ \bibinfo {pages} {1} (\bibinfo {year} {2018})},\ \Eprint
  {http://arxiv.org/abs/1712.06615} {arXiv:1712.06615 [astro-ph.CO]}
  \BibitemShut {NoStop}%
\bibitem [{\citenamefont {Cushman}\ \emph {et~al.}(2013)\citenamefont {Cushman}
  \emph {et~al.}}]{Cushman:2013zza}%
  \BibitemOpen
  \bibfield  {author} {\bibinfo {author} {\bibfnamefont {P.}~\bibnamefont
  {Cushman}} \emph {et~al.},\ }in\ \href@noop {} {\emph {\bibinfo {booktitle}
  {{Community Summer Study 2013}: {Snowmass on the Mississippi}}}}\ (\bibinfo
  {year} {2013})\ \Eprint {http://arxiv.org/abs/1310.8327} {arXiv:1310.8327
  [hep-ex]} \BibitemShut {NoStop}%
\bibitem [{\citenamefont {Schumann}(2019)}]{Schumann:2019eaa}%
  \BibitemOpen
  \bibfield  {author} {\bibinfo {author} {\bibfnamefont {M.}~\bibnamefont
  {Schumann}},\ }\href {\doibase 10.1088/1361-6471/ab2ea5} {\bibfield
  {journal} {\bibinfo  {journal} {J. Phys. G}\ }\textbf {\bibinfo {volume}
  {46}},\ \bibinfo {pages} {103003} (\bibinfo {year} {2019})},\ \Eprint
  {http://arxiv.org/abs/1903.03026} {arXiv:1903.03026 [astro-ph.CO]}
  \BibitemShut {NoStop}%
\bibitem [{\citenamefont {Akerib}\ \emph {et~al.}(2022)\citenamefont {Akerib}
  \emph {et~al.}}]{Akerib:2022ort}%
  \BibitemOpen
  \bibfield  {author} {\bibinfo {author} {\bibfnamefont {D.~S.}\ \bibnamefont
  {Akerib}} \emph {et~al.},\ }in\ \href@noop {} {\emph {\bibinfo {booktitle}
  {{2022 Snowmass Summer Study}}}}\ (\bibinfo {year} {2022})\ \Eprint
  {http://arxiv.org/abs/2203.08084} {arXiv:2203.08084 [hep-ex]} \BibitemShut
  {NoStop}%
\bibitem [{\citenamefont {Knapen}\ \emph {et~al.}(2017)\citenamefont {Knapen},
  \citenamefont {Lin},\ and\ \citenamefont {Zurek}}]{Knapen:2017xzo}%
  \BibitemOpen
  \bibfield  {author} {\bibinfo {author} {\bibfnamefont {S.}~\bibnamefont
  {Knapen}}, \bibinfo {author} {\bibfnamefont {T.}~\bibnamefont {Lin}}, \ and\
  \bibinfo {author} {\bibfnamefont {K.~M.}\ \bibnamefont {Zurek}},\ }\href
  {\doibase 10.1103/PhysRevD.96.115021} {\bibfield  {journal} {\bibinfo
  {journal} {Phys. Rev. D}\ }\textbf {\bibinfo {volume} {96}},\ \bibinfo
  {pages} {115021} (\bibinfo {year} {2017})},\ \Eprint
  {http://arxiv.org/abs/1709.07882} {arXiv:1709.07882 [hep-ph]} \BibitemShut
  {NoStop}%
\bibitem [{\citenamefont {Green}\ and\ \citenamefont
  {Kavanagh}(2021)}]{Green:2020jor}%
  \BibitemOpen
  \bibfield  {author} {\bibinfo {author} {\bibfnamefont {A.~M.}\ \bibnamefont
  {Green}}\ and\ \bibinfo {author} {\bibfnamefont {B.~J.}\ \bibnamefont
  {Kavanagh}},\ }\href {\doibase 10.1088/1361-6471/abc534} {\bibfield
  {journal} {\bibinfo  {journal} {J. Phys. G}\ }\textbf {\bibinfo {volume}
  {48}},\ \bibinfo {pages} {043001} (\bibinfo {year} {2021})},\ \Eprint
  {http://arxiv.org/abs/2007.10722} {arXiv:2007.10722 [astro-ph.CO]}
  \BibitemShut {NoStop}%
\bibitem [{\citenamefont {Adams}\ \emph {et~al.}(2022)\citenamefont {Adams}
  \emph {et~al.}}]{Adams:2022pbo}%
  \BibitemOpen
  \bibfield  {author} {\bibinfo {author} {\bibfnamefont {C.~B.}\ \bibnamefont
  {Adams}} \emph {et~al.},\ }in\ \href@noop {} {\emph {\bibinfo {booktitle}
  {{2022 Snowmass Summer Study}}}}\ (\bibinfo {year} {2022})\ \Eprint
  {http://arxiv.org/abs/2203.14923} {arXiv:2203.14923 [hep-ex]} \BibitemShut
  {NoStop}%
\bibitem [{\citenamefont {Carney}\ \emph {et~al.}(2022)\citenamefont {Carney}
  \emph {et~al.}}]{Carney:2022gse}%
  \BibitemOpen
  \bibfield  {author} {\bibinfo {author} {\bibfnamefont {D.}~\bibnamefont
  {Carney}} \emph {et~al.},\ }\href@noop {} {\  (\bibinfo {year} {2022})},\
  \Eprint {http://arxiv.org/abs/2203.06508} {arXiv:2203.06508 [hep-ph]}
  \BibitemShut {NoStop}%
\bibitem [{\citenamefont {Mitridate}\ \emph {et~al.}(2022)\citenamefont
  {Mitridate}, \citenamefont {Trickle}, \citenamefont {Zhang},\ and\
  \citenamefont {Zurek}}]{Mitridate:2022tnv}%
  \BibitemOpen
  \bibfield  {author} {\bibinfo {author} {\bibfnamefont {A.}~\bibnamefont
  {Mitridate}}, \bibinfo {author} {\bibfnamefont {T.}~\bibnamefont {Trickle}},
  \bibinfo {author} {\bibfnamefont {Z.}~\bibnamefont {Zhang}}, \ and\ \bibinfo
  {author} {\bibfnamefont {K.~M.}\ \bibnamefont {Zurek}},\ }in\ \href@noop {}
  {\emph {\bibinfo {booktitle} {{2022 Snowmass Summer Study}}}}\ (\bibinfo
  {year} {2022})\ \Eprint {http://arxiv.org/abs/2203.07492} {arXiv:2203.07492
  [hep-ph]} \BibitemShut {NoStop}%
\bibitem [{\citenamefont {Angloher}\ \emph {et~al.}(2016)\citenamefont
  {Angloher} \emph {et~al.}}]{CRESST:2015txj}%
  \BibitemOpen
  \bibfield  {author} {\bibinfo {author} {\bibfnamefont {G.}~\bibnamefont
  {Angloher}} \emph {et~al.} (\bibinfo {collaboration} {CRESST}),\ }\href
  {\doibase 10.1140/epjc/s10052-016-3877-3} {\bibfield  {journal} {\bibinfo
  {journal} {Eur. Phys. J. C}\ }\textbf {\bibinfo {volume} {76}},\ \bibinfo
  {pages} {25} (\bibinfo {year} {2016})},\ \Eprint
  {http://arxiv.org/abs/1509.01515} {arXiv:1509.01515 [astro-ph.CO]}
  \BibitemShut {NoStop}%
\bibitem [{\citenamefont {Angloher}\ \emph {et~al.}(2017)\citenamefont
  {Angloher} \emph {et~al.}}]{CRESST:2017ues}%
  \BibitemOpen
  \bibfield  {author} {\bibinfo {author} {\bibfnamefont {G.}~\bibnamefont
  {Angloher}} \emph {et~al.} (\bibinfo {collaboration} {CRESST}),\ }\href
  {\doibase 10.1140/epjc/s10052-017-5223-9} {\bibfield  {journal} {\bibinfo
  {journal} {Eur. Phys. J. C}\ }\textbf {\bibinfo {volume} {77}},\ \bibinfo
  {pages} {637} (\bibinfo {year} {2017})},\ \Eprint
  {http://arxiv.org/abs/1707.06749} {arXiv:1707.06749 [astro-ph.CO]}
  \BibitemShut {NoStop}%
\bibitem [{\citenamefont {Collar}(2018)}]{Collar:2018ydf}%
  \BibitemOpen
  \bibfield  {author} {\bibinfo {author} {\bibfnamefont {J.~I.}\ \bibnamefont
  {Collar}},\ }\href {\doibase 10.1103/PhysRevD.98.023005} {\bibfield
  {journal} {\bibinfo  {journal} {Phys. Rev. D}\ }\textbf {\bibinfo {volume}
  {98}},\ \bibinfo {pages} {023005} (\bibinfo {year} {2018})},\ \Eprint
  {http://arxiv.org/abs/1805.02646} {arXiv:1805.02646 [astro-ph.CO]}
  \BibitemShut {NoStop}%
\bibitem [{\citenamefont {Abdelhameed}\ \emph {et~al.}(2019)\citenamefont
  {Abdelhameed} \emph {et~al.}}]{CRESST:2019jnq}%
  \BibitemOpen
  \bibfield  {author} {\bibinfo {author} {\bibfnamefont {A.~H.}\ \bibnamefont
  {Abdelhameed}} \emph {et~al.} (\bibinfo {collaboration} {CRESST}),\ }\href
  {\doibase 10.1103/PhysRevD.100.102002} {\bibfield  {journal} {\bibinfo
  {journal} {Phys. Rev. D}\ }\textbf {\bibinfo {volume} {100}},\ \bibinfo
  {pages} {102002} (\bibinfo {year} {2019})},\ \Eprint
  {http://arxiv.org/abs/1904.00498} {arXiv:1904.00498 [astro-ph.CO]}
  \BibitemShut {NoStop}%
\bibitem [{\citenamefont {Armengaud}\ \emph {et~al.}(2019)\citenamefont
  {Armengaud} \emph {et~al.}}]{EDELWEISS:2019vjv}%
  \BibitemOpen
  \bibfield  {author} {\bibinfo {author} {\bibfnamefont {E.}~\bibnamefont
  {Armengaud}} \emph {et~al.} (\bibinfo {collaboration} {EDELWEISS}),\ }\href
  {\doibase 10.1103/PhysRevD.99.082003} {\bibfield  {journal} {\bibinfo
  {journal} {Phys. Rev. D}\ }\textbf {\bibinfo {volume} {99}},\ \bibinfo
  {pages} {082003} (\bibinfo {year} {2019})},\ \Eprint
  {http://arxiv.org/abs/1901.03588} {arXiv:1901.03588 [astro-ph.GA]}
  \BibitemShut {NoStop}%
\bibitem [{\citenamefont {Alkhatib}\ \emph {et~al.}(2021)\citenamefont
  {Alkhatib} \emph {et~al.}}]{SuperCDMS:2020aus}%
  \BibitemOpen
  \bibfield  {author} {\bibinfo {author} {\bibfnamefont {I.}~\bibnamefont
  {Alkhatib}} \emph {et~al.} (\bibinfo {collaboration} {SuperCDMS}),\ }\href
  {\doibase 10.1103/PhysRevLett.127.061801} {\bibfield  {journal} {\bibinfo
  {journal} {Phys. Rev. Lett.}\ }\textbf {\bibinfo {volume} {127}},\ \bibinfo
  {pages} {061801} (\bibinfo {year} {2021})},\ \Eprint
  {http://arxiv.org/abs/2007.14289} {arXiv:2007.14289 [hep-ex]} \BibitemShut
  {NoStop}%
\bibitem [{\citenamefont {Albakry}\ \emph
  {et~al.}(2022{\natexlab{a}})\citenamefont {Albakry} \emph
  {et~al.}}]{SuperCDMS:2022kse}%
  \BibitemOpen
  \bibfield  {author} {\bibinfo {author} {\bibfnamefont {M.~F.}\ \bibnamefont
  {Albakry}} \emph {et~al.} (\bibinfo {collaboration} {SuperCDMS}),\ }in\
  \href@noop {} {\emph {\bibinfo {booktitle} {{2022 Snowmass Summer Study}}}}\
  (\bibinfo {year} {2022})\ \Eprint {http://arxiv.org/abs/2203.08463}
  {arXiv:2203.08463 [physics.ins-det]} \BibitemShut {NoStop}%
\bibitem [{\citenamefont {Kouvaris}\ and\ \citenamefont
  {Pradler}(2017)}]{Kouvaris:2016afs}%
  \BibitemOpen
  \bibfield  {author} {\bibinfo {author} {\bibfnamefont {C.}~\bibnamefont
  {Kouvaris}}\ and\ \bibinfo {author} {\bibfnamefont {J.}~\bibnamefont
  {Pradler}},\ }\href {\doibase 10.1103/PhysRevLett.118.031803} {\bibfield
  {journal} {\bibinfo  {journal} {Phys. Rev. Lett.}\ }\textbf {\bibinfo
  {volume} {118}},\ \bibinfo {pages} {031803} (\bibinfo {year} {2017})},\
  \Eprint {http://arxiv.org/abs/1607.01789} {arXiv:1607.01789 [hep-ph]}
  \BibitemShut {NoStop}%
\bibitem [{\citenamefont {Ibe}\ \emph {et~al.}(2018)\citenamefont {Ibe},
  \citenamefont {Nakano}, \citenamefont {Shoji},\ and\ \citenamefont
  {Suzuki}}]{Ibe:2017yqa}%
  \BibitemOpen
  \bibfield  {author} {\bibinfo {author} {\bibfnamefont {M.}~\bibnamefont
  {Ibe}}, \bibinfo {author} {\bibfnamefont {W.}~\bibnamefont {Nakano}},
  \bibinfo {author} {\bibfnamefont {Y.}~\bibnamefont {Shoji}}, \ and\ \bibinfo
  {author} {\bibfnamefont {K.}~\bibnamefont {Suzuki}},\ }\href {\doibase
  10.1007/JHEP03(2018)194} {\bibfield  {journal} {\bibinfo  {journal} {JHEP}\
  }\textbf {\bibinfo {volume} {03}},\ \bibinfo {pages} {194} (\bibinfo {year}
  {2018})},\ \Eprint {http://arxiv.org/abs/1707.07258} {arXiv:1707.07258
  [hep-ph]} \BibitemShut {NoStop}%
\bibitem [{\citenamefont {Baxter}\ \emph {et~al.}(2020)\citenamefont {Baxter},
  \citenamefont {Kahn},\ and\ \citenamefont {Krnjaic}}]{Baxter:2019pnz}%
  \BibitemOpen
  \bibfield  {author} {\bibinfo {author} {\bibfnamefont {D.}~\bibnamefont
  {Baxter}}, \bibinfo {author} {\bibfnamefont {Y.}~\bibnamefont {Kahn}}, \ and\
  \bibinfo {author} {\bibfnamefont {G.}~\bibnamefont {Krnjaic}},\ }\href
  {\doibase 10.1103/PhysRevD.101.076014} {\bibfield  {journal} {\bibinfo
  {journal} {Phys. Rev. D}\ }\textbf {\bibinfo {volume} {101}},\ \bibinfo
  {pages} {076014} (\bibinfo {year} {2020})},\ \Eprint
  {http://arxiv.org/abs/1908.00012} {arXiv:1908.00012 [hep-ph]} \BibitemShut
  {NoStop}%
\bibitem [{\citenamefont {Bell}\ \emph {et~al.}(2020)\citenamefont {Bell},
  \citenamefont {Dent}, \citenamefont {Newstead}, \citenamefont {Sabharwal},\
  and\ \citenamefont {Weiler}}]{Bell:2019egg}%
  \BibitemOpen
  \bibfield  {author} {\bibinfo {author} {\bibfnamefont {N.~F.}\ \bibnamefont
  {Bell}}, \bibinfo {author} {\bibfnamefont {J.~B.}\ \bibnamefont {Dent}},
  \bibinfo {author} {\bibfnamefont {J.~L.}\ \bibnamefont {Newstead}}, \bibinfo
  {author} {\bibfnamefont {S.}~\bibnamefont {Sabharwal}}, \ and\ \bibinfo
  {author} {\bibfnamefont {T.~J.}\ \bibnamefont {Weiler}},\ }\href {\doibase
  10.1103/PhysRevD.101.015012} {\bibfield  {journal} {\bibinfo  {journal}
  {Phys. Rev. D}\ }\textbf {\bibinfo {volume} {101}},\ \bibinfo {pages}
  {015012} (\bibinfo {year} {2020})},\ \Eprint
  {http://arxiv.org/abs/1905.00046} {arXiv:1905.00046 [hep-ph]} \BibitemShut
  {NoStop}%
\bibitem [{\citenamefont {Essig}\ \emph {et~al.}(2020)\citenamefont {Essig},
  \citenamefont {Pradler}, \citenamefont {Sholapurkar},\ and\ \citenamefont
  {Yu}}]{Essig:2019xkx}%
  \BibitemOpen
  \bibfield  {author} {\bibinfo {author} {\bibfnamefont {R.}~\bibnamefont
  {Essig}}, \bibinfo {author} {\bibfnamefont {J.}~\bibnamefont {Pradler}},
  \bibinfo {author} {\bibfnamefont {M.}~\bibnamefont {Sholapurkar}}, \ and\
  \bibinfo {author} {\bibfnamefont {T.-T.}\ \bibnamefont {Yu}},\ }\href
  {\doibase 10.1103/PhysRevLett.124.021801} {\bibfield  {journal} {\bibinfo
  {journal} {Phys. Rev. Lett.}\ }\textbf {\bibinfo {volume} {124}},\ \bibinfo
  {pages} {021801} (\bibinfo {year} {2020})},\ \Eprint
  {http://arxiv.org/abs/1908.10881} {arXiv:1908.10881 [hep-ph]} \BibitemShut
  {NoStop}%
\bibitem [{\citenamefont {Aprile}\ \emph {et~al.}(2019)\citenamefont {Aprile}
  \emph {et~al.}}]{XENON:2019zpr}%
  \BibitemOpen
  \bibfield  {author} {\bibinfo {author} {\bibfnamefont {E.}~\bibnamefont
  {Aprile}} \emph {et~al.} (\bibinfo {collaboration} {XENON}),\ }\href
  {\doibase 10.1103/PhysRevLett.123.241803} {\bibfield  {journal} {\bibinfo
  {journal} {Phys. Rev. Lett.}\ }\textbf {\bibinfo {volume} {123}},\ \bibinfo
  {pages} {241803} (\bibinfo {year} {2019})},\ \Eprint
  {http://arxiv.org/abs/1907.12771} {arXiv:1907.12771 [hep-ex]} \BibitemShut
  {NoStop}%
\bibitem [{\citenamefont {Grilli~di Cortona}\ \emph {et~al.}(2020)\citenamefont
  {Grilli~di Cortona}, \citenamefont {Messina},\ and\ \citenamefont
  {Piacentini}}]{GrillidiCortona:2020owp}%
  \BibitemOpen
  \bibfield  {author} {\bibinfo {author} {\bibfnamefont {G.}~\bibnamefont
  {Grilli~di Cortona}}, \bibinfo {author} {\bibfnamefont {A.}~\bibnamefont
  {Messina}}, \ and\ \bibinfo {author} {\bibfnamefont {S.}~\bibnamefont
  {Piacentini}},\ }\href {\doibase 10.1007/JHEP11(2020)034} {\bibfield
  {journal} {\bibinfo  {journal} {JHEP}\ }\textbf {\bibinfo {volume} {11}},\
  \bibinfo {pages} {034} (\bibinfo {year} {2020})},\ \Eprint
  {http://arxiv.org/abs/2006.02453} {arXiv:2006.02453 [hep-ph]} \BibitemShut
  {NoStop}%
\bibitem [{\citenamefont {Knapen}\ \emph {et~al.}(2021)\citenamefont {Knapen},
  \citenamefont {Kozaczuk},\ and\ \citenamefont {Lin}}]{Knapen:2020aky}%
  \BibitemOpen
  \bibfield  {author} {\bibinfo {author} {\bibfnamefont {S.}~\bibnamefont
  {Knapen}}, \bibinfo {author} {\bibfnamefont {J.}~\bibnamefont {Kozaczuk}}, \
  and\ \bibinfo {author} {\bibfnamefont {T.}~\bibnamefont {Lin}},\ }\href
  {\doibase 10.1103/PhysRevLett.127.081805} {\bibfield  {journal} {\bibinfo
  {journal} {Phys. Rev. Lett.}\ }\textbf {\bibinfo {volume} {127}},\ \bibinfo
  {pages} {081805} (\bibinfo {year} {2021})},\ \Eprint
  {http://arxiv.org/abs/2011.09496} {arXiv:2011.09496 [hep-ph]} \BibitemShut
  {NoStop}%
\bibitem [{\citenamefont {{Armengaud}}\ \emph {et~al.}(2022)\citenamefont
  {{Armengaud}}, \citenamefont {{Arnaud}}, \citenamefont {{Augier}},
  \citenamefont {{Beno{\^\i}t}}, \citenamefont {{Berg{\'e}}}, \citenamefont
  {{Billard}}, \citenamefont {{Broniatowski}}, \citenamefont {{Camus}},
  \citenamefont {{Caze}}, \citenamefont {{Chapellier}}, \citenamefont
  {{Charlieux}}, \citenamefont {{De J{\'e}sus}}, \citenamefont {{Dumoulin}},
  \citenamefont {{Eitel}}, \citenamefont {{Filippini}}, \citenamefont
  {{Filosofov}}, \citenamefont {{Gascon}}, \citenamefont {{Giuliani}},
  \citenamefont {{Gros}}, \citenamefont {{Guy}}, \citenamefont {{Jin}},
  \citenamefont {{Juillard}}, \citenamefont {{Kleifges}}, \citenamefont
  {{Lattaud}}, \citenamefont {{Marnieros}}, \citenamefont {{Misiak}},
  \citenamefont {{Navick}}, \citenamefont {{Nones}}, \citenamefont
  {{Olivieri}}, \citenamefont {{Oriol}}, \citenamefont {{Pari}}, \citenamefont
  {{Paul}}, \citenamefont {{Poda}}, \citenamefont {{Rozov}}, \citenamefont
  {{Salagnac}}, \citenamefont {{Sanglard}}, \citenamefont {{Vagneron}},
  \citenamefont {{Yakushev}}, \citenamefont {{Zolotarova}},\ and\ \citenamefont
  {{Kavanagh}}}]{2022arXiv220303993A}%
  \BibitemOpen
  \bibfield  {author} {\bibinfo {author} {\bibfnamefont {E.}~\bibnamefont
  {{Armengaud}}}, \bibinfo {author} {\bibfnamefont {Q.}~\bibnamefont
  {{Arnaud}}}, \bibinfo {author} {\bibfnamefont {C.}~\bibnamefont {{Augier}}},
  \bibinfo {author} {\bibfnamefont {A.}~\bibnamefont {{Beno{\^\i}t}}}, \bibinfo
  {author} {\bibfnamefont {L.}~\bibnamefont {{Berg{\'e}}}}, \bibinfo {author}
  {\bibfnamefont {J.}~\bibnamefont {{Billard}}}, \bibinfo {author}
  {\bibfnamefont {A.}~\bibnamefont {{Broniatowski}}}, \bibinfo {author}
  {\bibfnamefont {P.}~\bibnamefont {{Camus}}}, \bibinfo {author} {\bibfnamefont
  {A.}~\bibnamefont {{Caze}}}, \bibinfo {author} {\bibfnamefont
  {M.}~\bibnamefont {{Chapellier}}}, \bibinfo {author} {\bibfnamefont
  {F.}~\bibnamefont {{Charlieux}}}, \bibinfo {author} {\bibfnamefont
  {M.}~\bibnamefont {{De J{\'e}sus}}}, \bibinfo {author} {\bibfnamefont
  {L.}~\bibnamefont {{Dumoulin}}}, \bibinfo {author} {\bibfnamefont
  {K.}~\bibnamefont {{Eitel}}}, \bibinfo {author} {\bibfnamefont {J.~B.}\
  \bibnamefont {{Filippini}}}, \bibinfo {author} {\bibfnamefont
  {D.}~\bibnamefont {{Filosofov}}}, \bibinfo {author} {\bibfnamefont
  {J.}~\bibnamefont {{Gascon}}}, \bibinfo {author} {\bibfnamefont
  {A.}~\bibnamefont {{Giuliani}}}, \bibinfo {author} {\bibfnamefont
  {M.}~\bibnamefont {{Gros}}}, \bibinfo {author} {\bibfnamefont
  {E.}~\bibnamefont {{Guy}}}, \bibinfo {author} {\bibfnamefont
  {Y.}~\bibnamefont {{Jin}}}, \bibinfo {author} {\bibfnamefont
  {A.}~\bibnamefont {{Juillard}}}, \bibinfo {author} {\bibfnamefont
  {M.}~\bibnamefont {{Kleifges}}}, \bibinfo {author} {\bibfnamefont
  {H.}~\bibnamefont {{Lattaud}}}, \bibinfo {author} {\bibfnamefont
  {S.}~\bibnamefont {{Marnieros}}}, \bibinfo {author} {\bibfnamefont
  {D.}~\bibnamefont {{Misiak}}}, \bibinfo {author} {\bibfnamefont {X.~F.}\
  \bibnamefont {{Navick}}}, \bibinfo {author} {\bibfnamefont {C.}~\bibnamefont
  {{Nones}}}, \bibinfo {author} {\bibfnamefont {E.}~\bibnamefont {{Olivieri}}},
  \bibinfo {author} {\bibfnamefont {C.}~\bibnamefont {{Oriol}}}, \bibinfo
  {author} {\bibfnamefont {P.}~\bibnamefont {{Pari}}}, \bibinfo {author}
  {\bibfnamefont {B.}~\bibnamefont {{Paul}}}, \bibinfo {author} {\bibfnamefont
  {D.}~\bibnamefont {{Poda}}}, \bibinfo {author} {\bibfnamefont
  {S.}~\bibnamefont {{Rozov}}}, \bibinfo {author} {\bibfnamefont
  {T.}~\bibnamefont {{Salagnac}}}, \bibinfo {author} {\bibfnamefont
  {V.}~\bibnamefont {{Sanglard}}}, \bibinfo {author} {\bibfnamefont
  {L.}~\bibnamefont {{Vagneron}}}, \bibinfo {author} {\bibfnamefont
  {E.}~\bibnamefont {{Yakushev}}}, \bibinfo {author} {\bibfnamefont
  {A.}~\bibnamefont {{Zolotarova}}}, \ and\ \bibinfo {author} {\bibfnamefont
  {B.~J.}\ \bibnamefont {{Kavanagh}}},\ }\href@noop {} {\bibfield  {journal}
  {\bibinfo  {journal} {arXiv e-prints}\ ,\ \bibinfo {eid} {arXiv:2203.03993}}
  (\bibinfo {year} {2022})},\ \Eprint {http://arxiv.org/abs/2203.03993}
  {arXiv:2203.03993 [astro-ph.GA]} \BibitemShut {NoStop}%
\bibitem [{\citenamefont {Bringmann}\ and\ \citenamefont
  {Pospelov}(2019)}]{Bringmann:2018cvk}%
  \BibitemOpen
  \bibfield  {author} {\bibinfo {author} {\bibfnamefont {T.}~\bibnamefont
  {Bringmann}}\ and\ \bibinfo {author} {\bibfnamefont {M.}~\bibnamefont
  {Pospelov}},\ }\href {\doibase 10.1103/PhysRevLett.122.171801} {\bibfield
  {journal} {\bibinfo  {journal} {Phys. Rev. Lett.}\ }\textbf {\bibinfo
  {volume} {122}},\ \bibinfo {pages} {171801} (\bibinfo {year} {2019})},\
  \Eprint {http://arxiv.org/abs/1810.10543} {arXiv:1810.10543 [hep-ph]}
  \BibitemShut {NoStop}%
\bibitem [{\citenamefont {Ema}\ \emph {et~al.}(2019)\citenamefont {Ema},
  \citenamefont {Sala},\ and\ \citenamefont {Sato}}]{Ema:2018bih}%
  \BibitemOpen
  \bibfield  {author} {\bibinfo {author} {\bibfnamefont {Y.}~\bibnamefont
  {Ema}}, \bibinfo {author} {\bibfnamefont {F.}~\bibnamefont {Sala}}, \ and\
  \bibinfo {author} {\bibfnamefont {R.}~\bibnamefont {Sato}},\ }\href {\doibase
  10.1103/PhysRevLett.122.181802} {\bibfield  {journal} {\bibinfo  {journal}
  {Phys. Rev. Lett.}\ }\textbf {\bibinfo {volume} {122}},\ \bibinfo {pages}
  {181802} (\bibinfo {year} {2019})},\ \Eprint
  {http://arxiv.org/abs/1811.00520} {arXiv:1811.00520 [hep-ph]} \BibitemShut
  {NoStop}%
\bibitem [{\citenamefont {Bondarenko}\ \emph {et~al.}(2020)\citenamefont
  {Bondarenko}, \citenamefont {Boyarsky}, \citenamefont {Bringmann},
  \citenamefont {Hufnagel}, \citenamefont {Schmidt-Hoberg},\ and\ \citenamefont
  {Sokolenko}}]{Bondarenko:2019vrb}%
  \BibitemOpen
  \bibfield  {author} {\bibinfo {author} {\bibfnamefont {K.}~\bibnamefont
  {Bondarenko}}, \bibinfo {author} {\bibfnamefont {A.}~\bibnamefont
  {Boyarsky}}, \bibinfo {author} {\bibfnamefont {T.}~\bibnamefont {Bringmann}},
  \bibinfo {author} {\bibfnamefont {M.}~\bibnamefont {Hufnagel}}, \bibinfo
  {author} {\bibfnamefont {K.}~\bibnamefont {Schmidt-Hoberg}}, \ and\ \bibinfo
  {author} {\bibfnamefont {A.}~\bibnamefont {Sokolenko}},\ }\href {\doibase
  10.1007/JHEP03(2020)118} {\bibfield  {journal} {\bibinfo  {journal} {JHEP}\
  }\textbf {\bibinfo {volume} {03}},\ \bibinfo {pages} {118} (\bibinfo {year}
  {2020})},\ \Eprint {http://arxiv.org/abs/1909.08632} {arXiv:1909.08632
  [hep-ph]} \BibitemShut {NoStop}%
\bibitem [{\citenamefont {Cappiello}\ and\ \citenamefont
  {Beacom}(2019)}]{Cappiello:2019qsw}%
  \BibitemOpen
  \bibfield  {author} {\bibinfo {author} {\bibfnamefont {C.~V.}\ \bibnamefont
  {Cappiello}}\ and\ \bibinfo {author} {\bibfnamefont {J.~F.}\ \bibnamefont
  {Beacom}},\ }\href {\doibase 10.1103/PhysRevD.104.069901} {\bibfield
  {journal} {\bibinfo  {journal} {Phys. Rev. D}\ }\textbf {\bibinfo {volume}
  {100}},\ \bibinfo {pages} {103011} (\bibinfo {year} {2019})},\ \bibinfo
  {note} {[Erratum: Phys.Rev.D 104, 069901(E) (2021)]},\ \Eprint
  {http://arxiv.org/abs/1906.11283} {arXiv:1906.11283 [hep-ph]} \BibitemShut
  {NoStop}%
\bibitem [{\citenamefont {Dent}\ \emph {et~al.}(2020)\citenamefont {Dent},
  \citenamefont {Dutta}, \citenamefont {Newstead},\ and\ \citenamefont
  {Shoemaker}}]{Dent:2019krz}%
  \BibitemOpen
  \bibfield  {author} {\bibinfo {author} {\bibfnamefont {J.~B.}\ \bibnamefont
  {Dent}}, \bibinfo {author} {\bibfnamefont {B.}~\bibnamefont {Dutta}},
  \bibinfo {author} {\bibfnamefont {J.~L.}\ \bibnamefont {Newstead}}, \ and\
  \bibinfo {author} {\bibfnamefont {I.~M.}\ \bibnamefont {Shoemaker}},\ }\href
  {\doibase 10.1103/PhysRevD.101.116007} {\bibfield  {journal} {\bibinfo
  {journal} {Phys. Rev. D}\ }\textbf {\bibinfo {volume} {101}},\ \bibinfo
  {pages} {116007} (\bibinfo {year} {2020})},\ \Eprint
  {http://arxiv.org/abs/1907.03782} {arXiv:1907.03782 [hep-ph]} \BibitemShut
  {NoStop}%
\bibitem [{\citenamefont {Andriamirado}\ \emph {et~al.}(2021)\citenamefont
  {Andriamirado} \emph {et~al.}}]{PROSPECT:2021awi}%
  \BibitemOpen
  \bibfield  {author} {\bibinfo {author} {\bibfnamefont {M.}~\bibnamefont
  {Andriamirado}} \emph {et~al.} (\bibinfo {collaboration} {PROSPECT, (PROSPECT
  Collaboration)*}),\ }\href {\doibase 10.1103/PhysRevD.104.012009} {\bibfield
  {journal} {\bibinfo  {journal} {Phys. Rev. D}\ }\textbf {\bibinfo {volume}
  {104}},\ \bibinfo {pages} {012009} (\bibinfo {year} {2021})},\ \Eprint
  {http://arxiv.org/abs/2104.11219} {arXiv:2104.11219 [hep-ex]} \BibitemShut
  {NoStop}%
\bibitem [{\citenamefont {Xu}\ \emph {et~al.}(2022)\citenamefont {Xu} \emph
  {et~al.}}]{CDEX:2022fig}%
  \BibitemOpen
  \bibfield  {author} {\bibinfo {author} {\bibfnamefont {R.}~\bibnamefont {Xu}}
  \emph {et~al.} (\bibinfo {collaboration} {CDEX}),\ }\href {\doibase
  10.1103/PhysRevD.106.052008} {\bibfield  {journal} {\bibinfo  {journal}
  {Phys. Rev. D}\ }\textbf {\bibinfo {volume} {106}},\ \bibinfo {pages}
  {052008} (\bibinfo {year} {2022})},\ \Eprint
  {http://arxiv.org/abs/2201.01704} {arXiv:2201.01704 [hep-ex]} \BibitemShut
  {NoStop}%
\bibitem [{\citenamefont {Maity}\ and\ \citenamefont
  {Laha}(2022)}]{Maity:2022exk}%
  \BibitemOpen
  \bibfield  {author} {\bibinfo {author} {\bibfnamefont {T.~N.}\ \bibnamefont
  {Maity}}\ and\ \bibinfo {author} {\bibfnamefont {R.}~\bibnamefont {Laha}},\
  }\href@noop {} {\  (\bibinfo {year} {2022})},\ \Eprint
  {http://arxiv.org/abs/2210.01815} {arXiv:2210.01815 [hep-ph]} \BibitemShut
  {NoStop}%
\bibitem [{\citenamefont {Abe}\ \emph {et~al.}(2022)\citenamefont {Abe} \emph
  {et~al.}}]{Super-Kamiokande:2022ncz}%
  \BibitemOpen
  \bibfield  {author} {\bibinfo {author} {\bibfnamefont {K.}~\bibnamefont
  {Abe}} \emph {et~al.} (\bibinfo {collaboration} {Super-Kamiokande}),\
  }\href@noop {} {\  (\bibinfo {year} {2022})},\ \Eprint
  {http://arxiv.org/abs/2209.14968} {arXiv:2209.14968 [hep-ex]} \BibitemShut
  {NoStop}%
\bibitem [{\citenamefont {Zhang}(2022)}]{Zhang:2020nis}%
  \BibitemOpen
  \bibfield  {author} {\bibinfo {author} {\bibfnamefont {Y.}~\bibnamefont
  {Zhang}},\ }\href {\doibase 10.1093/ptep/ptab156} {\bibfield  {journal}
  {\bibinfo  {journal} {PTEP}\ }\textbf {\bibinfo {volume} {2022}},\ \bibinfo
  {pages} {013B05} (\bibinfo {year} {2022})},\ \Eprint
  {http://arxiv.org/abs/2001.00948} {arXiv:2001.00948 [hep-ph]} \BibitemShut
  {NoStop}%
\bibitem [{\citenamefont {Das}\ and\ \citenamefont {Sen}(2021)}]{Das:2021lcr}%
  \BibitemOpen
  \bibfield  {author} {\bibinfo {author} {\bibfnamefont {A.}~\bibnamefont
  {Das}}\ and\ \bibinfo {author} {\bibfnamefont {M.}~\bibnamefont {Sen}},\
  }\href {\doibase 10.1103/PhysRevD.104.075029} {\bibfield  {journal} {\bibinfo
   {journal} {Phys. Rev. D}\ }\textbf {\bibinfo {volume} {104}},\ \bibinfo
  {pages} {075029} (\bibinfo {year} {2021})},\ \Eprint
  {http://arxiv.org/abs/2104.00027} {arXiv:2104.00027 [hep-ph]} \BibitemShut
  {NoStop}%
\bibitem [{\citenamefont {Lin}\ \emph {et~al.}(2022)\citenamefont {Lin},
  \citenamefont {Wu}, \citenamefont {Wu},\ and\ \citenamefont
  {Wong}}]{Lin:2022dbl}%
  \BibitemOpen
  \bibfield  {author} {\bibinfo {author} {\bibfnamefont {Y.-H.}\ \bibnamefont
  {Lin}}, \bibinfo {author} {\bibfnamefont {W.-H.}\ \bibnamefont {Wu}},
  \bibinfo {author} {\bibfnamefont {M.-R.}\ \bibnamefont {Wu}}, \ and\ \bibinfo
  {author} {\bibfnamefont {H.~T.-K.}\ \bibnamefont {Wong}},\ }\href@noop {} {\
  (\bibinfo {year} {2022})},\ \Eprint {http://arxiv.org/abs/2206.06864}
  {arXiv:2206.06864 [hep-ph]} \BibitemShut {NoStop}%
\bibitem [{\citenamefont {Carenza}\ and\ \citenamefont {De~la
  Torre~Luque}(2022)}]{Carenza:2022som}%
  \BibitemOpen
  \bibfield  {author} {\bibinfo {author} {\bibfnamefont {P.}~\bibnamefont
  {Carenza}}\ and\ \bibinfo {author} {\bibfnamefont {P.}~\bibnamefont {De~la
  Torre~Luque}},\ }\href@noop {} {\  (\bibinfo {year} {2022})},\ \Eprint
  {http://arxiv.org/abs/2210.17206} {arXiv:2210.17206 [astro-ph.HE]}
  \BibitemShut {NoStop}%
\bibitem [{\citenamefont {Wang}\ \emph {et~al.}(2022)\citenamefont {Wang},
  \citenamefont {Granelli},\ and\ \citenamefont {Ullio}}]{Wang:2021jic}%
  \BibitemOpen
  \bibfield  {author} {\bibinfo {author} {\bibfnamefont {J.-W.}\ \bibnamefont
  {Wang}}, \bibinfo {author} {\bibfnamefont {A.}~\bibnamefont {Granelli}}, \
  and\ \bibinfo {author} {\bibfnamefont {P.}~\bibnamefont {Ullio}},\ }\href
  {\doibase 10.1103/PhysRevLett.128.221104} {\bibfield  {journal} {\bibinfo
  {journal} {Phys. Rev. Lett.}\ }\textbf {\bibinfo {volume} {128}},\ \bibinfo
  {pages} {221104} (\bibinfo {year} {2022})},\ \Eprint
  {http://arxiv.org/abs/2111.13644} {arXiv:2111.13644 [astro-ph.HE]}
  \BibitemShut {NoStop}%
\bibitem [{\citenamefont {Granelli}\ \emph {et~al.}(2022)\citenamefont
  {Granelli}, \citenamefont {Ullio},\ and\ \citenamefont
  {Wang}}]{Granelli:2022ysi}%
  \BibitemOpen
  \bibfield  {author} {\bibinfo {author} {\bibfnamefont {A.}~\bibnamefont
  {Granelli}}, \bibinfo {author} {\bibfnamefont {P.}~\bibnamefont {Ullio}}, \
  and\ \bibinfo {author} {\bibfnamefont {J.-W.}\ \bibnamefont {Wang}},\ }\href
  {\doibase 10.1088/1475-7516/2022/07/013} {\bibfield  {journal} {\bibinfo
  {journal} {JCAP}\ }\textbf {\bibinfo {volume} {07}},\ \bibinfo {pages} {013}
  (\bibinfo {year} {2022})},\ \Eprint {http://arxiv.org/abs/2202.07598}
  {arXiv:2202.07598 [astro-ph.HE]} \BibitemShut {NoStop}%
\bibitem [{\citenamefont {An}\ \emph {et~al.}(2018)\citenamefont {An},
  \citenamefont {Pospelov}, \citenamefont {Pradler},\ and\ \citenamefont
  {Ritz}}]{An:2017ojc}%
  \BibitemOpen
  \bibfield  {author} {\bibinfo {author} {\bibfnamefont {H.}~\bibnamefont
  {An}}, \bibinfo {author} {\bibfnamefont {M.}~\bibnamefont {Pospelov}},
  \bibinfo {author} {\bibfnamefont {J.}~\bibnamefont {Pradler}}, \ and\
  \bibinfo {author} {\bibfnamefont {A.}~\bibnamefont {Ritz}},\ }\href {\doibase
  10.1103/PhysRevLett.120.141801} {\bibfield  {journal} {\bibinfo  {journal}
  {Phys. Rev. Lett.}\ }\textbf {\bibinfo {volume} {120}},\ \bibinfo {pages}
  {141801} (\bibinfo {year} {2018})},\ \bibinfo {note} {[Erratum:
  Phys.Rev.Lett. 121, 259903(E) (2018)]},\ \Eprint
  {http://arxiv.org/abs/1708.03642} {arXiv:1708.03642 [hep-ph]} \BibitemShut
  {NoStop}%
\bibitem [{\citenamefont {Emken}\ \emph {et~al.}(2018)\citenamefont {Emken},
  \citenamefont {Kouvaris},\ and\ \citenamefont {Nielsen}}]{Emken:2017hnp}%
  \BibitemOpen
  \bibfield  {author} {\bibinfo {author} {\bibfnamefont {T.}~\bibnamefont
  {Emken}}, \bibinfo {author} {\bibfnamefont {C.}~\bibnamefont {Kouvaris}}, \
  and\ \bibinfo {author} {\bibfnamefont {N.~G.}\ \bibnamefont {Nielsen}},\
  }\href {\doibase 10.1103/PhysRevD.97.063007} {\bibfield  {journal} {\bibinfo
  {journal} {Phys. Rev. D}\ }\textbf {\bibinfo {volume} {97}},\ \bibinfo
  {pages} {063007} (\bibinfo {year} {2018})},\ \Eprint
  {http://arxiv.org/abs/1709.06573} {arXiv:1709.06573 [hep-ph]} \BibitemShut
  {NoStop}%
\bibitem [{\citenamefont {Agashe}\ \emph {et~al.}(2014)\citenamefont {Agashe},
  \citenamefont {Cui}, \citenamefont {Necib},\ and\ \citenamefont
  {Thaler}}]{Agashe:2014yua}%
  \BibitemOpen
  \bibfield  {author} {\bibinfo {author} {\bibfnamefont {K.}~\bibnamefont
  {Agashe}}, \bibinfo {author} {\bibfnamefont {Y.}~\bibnamefont {Cui}},
  \bibinfo {author} {\bibfnamefont {L.}~\bibnamefont {Necib}}, \ and\ \bibinfo
  {author} {\bibfnamefont {J.}~\bibnamefont {Thaler}},\ }\href {\doibase
  10.1088/1475-7516/2014/10/062} {\bibfield  {journal} {\bibinfo  {journal}
  {JCAP}\ }\textbf {\bibinfo {volume} {10}},\ \bibinfo {pages} {062} (\bibinfo
  {year} {2014})},\ \Eprint {http://arxiv.org/abs/1405.7370} {arXiv:1405.7370
  [hep-ph]} \BibitemShut {NoStop}%
\bibitem [{\citenamefont {Berger}\ \emph {et~al.}(2015)\citenamefont {Berger},
  \citenamefont {Cui},\ and\ \citenamefont {Zhao}}]{Berger:2014sqa}%
  \BibitemOpen
  \bibfield  {author} {\bibinfo {author} {\bibfnamefont {J.}~\bibnamefont
  {Berger}}, \bibinfo {author} {\bibfnamefont {Y.}~\bibnamefont {Cui}}, \ and\
  \bibinfo {author} {\bibfnamefont {Y.}~\bibnamefont {Zhao}},\ }\href {\doibase
  10.1088/1475-7516/2015/02/005} {\bibfield  {journal} {\bibinfo  {journal}
  {JCAP}\ }\textbf {\bibinfo {volume} {02}},\ \bibinfo {pages} {005} (\bibinfo
  {year} {2015})},\ \Eprint {http://arxiv.org/abs/1410.2246} {arXiv:1410.2246
  [hep-ph]} \BibitemShut {NoStop}%
\bibitem [{\citenamefont {Kong}\ \emph {et~al.}(2015)\citenamefont {Kong},
  \citenamefont {Mohlabeng},\ and\ \citenamefont {Park}}]{Kong:2014mia}%
  \BibitemOpen
  \bibfield  {author} {\bibinfo {author} {\bibfnamefont {K.}~\bibnamefont
  {Kong}}, \bibinfo {author} {\bibfnamefont {G.}~\bibnamefont {Mohlabeng}}, \
  and\ \bibinfo {author} {\bibfnamefont {J.-C.}\ \bibnamefont {Park}},\ }\href
  {\doibase 10.1016/j.physletb.2015.02.057} {\bibfield  {journal} {\bibinfo
  {journal} {Phys. Lett. B}\ }\textbf {\bibinfo {volume} {743}},\ \bibinfo
  {pages} {256} (\bibinfo {year} {2015})},\ \Eprint
  {http://arxiv.org/abs/1411.6632} {arXiv:1411.6632 [hep-ph]} \BibitemShut
  {NoStop}%
\bibitem [{\citenamefont {Cherry}\ \emph {et~al.}(2015)\citenamefont {Cherry},
  \citenamefont {Frandsen},\ and\ \citenamefont {Shoemaker}}]{Cherry:2015oca}%
  \BibitemOpen
  \bibfield  {author} {\bibinfo {author} {\bibfnamefont {J.~F.}\ \bibnamefont
  {Cherry}}, \bibinfo {author} {\bibfnamefont {M.~T.}\ \bibnamefont
  {Frandsen}}, \ and\ \bibinfo {author} {\bibfnamefont {I.~M.}\ \bibnamefont
  {Shoemaker}},\ }\href {\doibase 10.1103/PhysRevLett.114.231303} {\bibfield
  {journal} {\bibinfo  {journal} {Phys. Rev. Lett.}\ }\textbf {\bibinfo
  {volume} {114}},\ \bibinfo {pages} {231303} (\bibinfo {year} {2015})},\
  \Eprint {http://arxiv.org/abs/1501.03166} {arXiv:1501.03166 [hep-ph]}
  \BibitemShut {NoStop}%
\bibitem [{\citenamefont {Kopp}\ \emph {et~al.}(2015)\citenamefont {Kopp},
  \citenamefont {Liu},\ and\ \citenamefont {Wang}}]{Kopp:2015bfa}%
  \BibitemOpen
  \bibfield  {author} {\bibinfo {author} {\bibfnamefont {J.}~\bibnamefont
  {Kopp}}, \bibinfo {author} {\bibfnamefont {J.}~\bibnamefont {Liu}}, \ and\
  \bibinfo {author} {\bibfnamefont {X.-P.}\ \bibnamefont {Wang}},\ }\href
  {\doibase 10.1007/JHEP04(2015)105} {\bibfield  {journal} {\bibinfo  {journal}
  {JHEP}\ }\textbf {\bibinfo {volume} {04}},\ \bibinfo {pages} {105} (\bibinfo
  {year} {2015})},\ \Eprint {http://arxiv.org/abs/1503.02669} {arXiv:1503.02669
  [hep-ph]} \BibitemShut {NoStop}%
\bibitem [{\citenamefont {Alhazmi}\ \emph {et~al.}(2017)\citenamefont
  {Alhazmi}, \citenamefont {Kong}, \citenamefont {Mohlabeng},\ and\
  \citenamefont {Park}}]{Alhazmi:2016qcs}%
  \BibitemOpen
  \bibfield  {author} {\bibinfo {author} {\bibfnamefont {H.}~\bibnamefont
  {Alhazmi}}, \bibinfo {author} {\bibfnamefont {K.}~\bibnamefont {Kong}},
  \bibinfo {author} {\bibfnamefont {G.}~\bibnamefont {Mohlabeng}}, \ and\
  \bibinfo {author} {\bibfnamefont {J.-C.}\ \bibnamefont {Park}},\ }\href
  {\doibase 10.1007/JHEP04(2017)158} {\bibfield  {journal} {\bibinfo  {journal}
  {JHEP}\ }\textbf {\bibinfo {volume} {04}},\ \bibinfo {pages} {158} (\bibinfo
  {year} {2017})},\ \Eprint {http://arxiv.org/abs/1611.09866} {arXiv:1611.09866
  [hep-ph]} \BibitemShut {NoStop}%
\bibitem [{\citenamefont {Bhattacharya}\ \emph {et~al.}(2017)\citenamefont
  {Bhattacharya}, \citenamefont {Gandhi}, \citenamefont {Gupta},\ and\
  \citenamefont {Mukhopadhyay}}]{Bhattacharya:2016tma}%
  \BibitemOpen
  \bibfield  {author} {\bibinfo {author} {\bibfnamefont {A.}~\bibnamefont
  {Bhattacharya}}, \bibinfo {author} {\bibfnamefont {R.}~\bibnamefont
  {Gandhi}}, \bibinfo {author} {\bibfnamefont {A.}~\bibnamefont {Gupta}}, \
  and\ \bibinfo {author} {\bibfnamefont {S.}~\bibnamefont {Mukhopadhyay}},\
  }\href {\doibase 10.1088/1475-7516/2017/05/002} {\bibfield  {journal}
  {\bibinfo  {journal} {JCAP}\ }\textbf {\bibinfo {volume} {05}},\ \bibinfo
  {pages} {002} (\bibinfo {year} {2017})},\ \Eprint
  {http://arxiv.org/abs/1612.02834} {arXiv:1612.02834 [hep-ph]} \BibitemShut
  {NoStop}%
\bibitem [{\citenamefont {Necib}\ \emph {et~al.}(2017)\citenamefont {Necib},
  \citenamefont {Moon}, \citenamefont {Wongjirad},\ and\ \citenamefont
  {Conrad}}]{Necib:2016aez}%
  \BibitemOpen
  \bibfield  {author} {\bibinfo {author} {\bibfnamefont {L.}~\bibnamefont
  {Necib}}, \bibinfo {author} {\bibfnamefont {J.}~\bibnamefont {Moon}},
  \bibinfo {author} {\bibfnamefont {T.}~\bibnamefont {Wongjirad}}, \ and\
  \bibinfo {author} {\bibfnamefont {J.~M.}\ \bibnamefont {Conrad}},\ }\href
  {\doibase 10.1103/PhysRevD.95.075018} {\bibfield  {journal} {\bibinfo
  {journal} {Phys. Rev. D}\ }\textbf {\bibinfo {volume} {95}},\ \bibinfo
  {pages} {075018} (\bibinfo {year} {2017})},\ \Eprint
  {http://arxiv.org/abs/1610.03486} {arXiv:1610.03486 [hep-ph]} \BibitemShut
  {NoStop}%
\bibitem [{\citenamefont {Kachulis}\ \emph {et~al.}(2018)\citenamefont
  {Kachulis} \emph {et~al.}}]{Super-Kamiokande:2017dch}%
  \BibitemOpen
  \bibfield  {author} {\bibinfo {author} {\bibfnamefont {C.}~\bibnamefont
  {Kachulis}} \emph {et~al.} (\bibinfo {collaboration} {Super-Kamiokande}),\
  }\href {\doibase 10.1103/PhysRevLett.120.221301} {\bibfield  {journal}
  {\bibinfo  {journal} {Phys. Rev. Lett.}\ }\textbf {\bibinfo {volume} {120}},\
  \bibinfo {pages} {221301} (\bibinfo {year} {2018})},\ \Eprint
  {http://arxiv.org/abs/1711.05278} {arXiv:1711.05278 [hep-ex]} \BibitemShut
  {NoStop}%
\bibitem [{\citenamefont {Marfatia}\ and\ \citenamefont
  {Tseng}(2022)}]{Marfatia:2022jiz}%
  \BibitemOpen
  \bibfield  {author} {\bibinfo {author} {\bibfnamefont {D.}~\bibnamefont
  {Marfatia}}\ and\ \bibinfo {author} {\bibfnamefont {P.-Y.}\ \bibnamefont
  {Tseng}},\ }\href@noop {} {\  (\bibinfo {year} {2022})},\ \Eprint
  {http://arxiv.org/abs/2212.13035} {arXiv:2212.13035 [hep-ph]} \BibitemShut
  {NoStop}%
\bibitem [{\citenamefont {Hu}\ \emph {et~al.}(2017)\citenamefont {Hu},
  \citenamefont {Kusenko},\ and\ \citenamefont {Takhistov}}]{Hu:2016xas}%
  \BibitemOpen
  \bibfield  {author} {\bibinfo {author} {\bibfnamefont {P.-K.}\ \bibnamefont
  {Hu}}, \bibinfo {author} {\bibfnamefont {A.}~\bibnamefont {Kusenko}}, \ and\
  \bibinfo {author} {\bibfnamefont {V.}~\bibnamefont {Takhistov}},\ }\href
  {\doibase 10.1016/j.physletb.2017.02.035} {\bibfield  {journal} {\bibinfo
  {journal} {Phys. Lett. B}\ }\textbf {\bibinfo {volume} {768}},\ \bibinfo
  {pages} {18} (\bibinfo {year} {2017})},\ \Eprint
  {http://arxiv.org/abs/1611.04599} {arXiv:1611.04599 [hep-ph]} \BibitemShut
  {NoStop}%
\bibitem [{\citenamefont {Dunsky}\ \emph {et~al.}(2019)\citenamefont {Dunsky},
  \citenamefont {Hall},\ and\ \citenamefont {Harigaya}}]{Dunsky:2018mqs}%
  \BibitemOpen
  \bibfield  {author} {\bibinfo {author} {\bibfnamefont {D.}~\bibnamefont
  {Dunsky}}, \bibinfo {author} {\bibfnamefont {L.~J.}\ \bibnamefont {Hall}}, \
  and\ \bibinfo {author} {\bibfnamefont {K.}~\bibnamefont {Harigaya}},\ }\href
  {\doibase 10.1088/1475-7516/2019/07/015} {\bibfield  {journal} {\bibinfo
  {journal} {JCAP}\ }\textbf {\bibinfo {volume} {07}},\ \bibinfo {pages} {015}
  (\bibinfo {year} {2019})},\ \Eprint {http://arxiv.org/abs/1812.11116}
  {arXiv:1812.11116 [astro-ph.HE]} \BibitemShut {NoStop}%
\bibitem [{\citenamefont {Li}\ and\ \citenamefont {Lin}(2020)}]{Li:2020wyl}%
  \BibitemOpen
  \bibfield  {author} {\bibinfo {author} {\bibfnamefont {J.-T.}\ \bibnamefont
  {Li}}\ and\ \bibinfo {author} {\bibfnamefont {T.}~\bibnamefont {Lin}},\
  }\href {\doibase 10.1103/PhysRevD.101.103034} {\bibfield  {journal} {\bibinfo
   {journal} {Phys. Rev. D}\ }\textbf {\bibinfo {volume} {101}},\ \bibinfo
  {pages} {103034} (\bibinfo {year} {2020})},\ \Eprint
  {http://arxiv.org/abs/2002.04625} {arXiv:2002.04625 [astro-ph.CO]}
  \BibitemShut {NoStop}%
\bibitem [{\citenamefont {Besla}\ \emph {et~al.}(2019)\citenamefont {Besla},
  \citenamefont {Peter},\ and\ \citenamefont
  {Garavito-Camargo}}]{Besla:2019xbx}%
  \BibitemOpen
  \bibfield  {author} {\bibinfo {author} {\bibfnamefont {G.}~\bibnamefont
  {Besla}}, \bibinfo {author} {\bibfnamefont {A.}~\bibnamefont {Peter}}, \ and\
  \bibinfo {author} {\bibfnamefont {N.}~\bibnamefont {Garavito-Camargo}},\
  }\href {\doibase 10.1088/1475-7516/2019/11/013} {\bibfield  {journal}
  {\bibinfo  {journal} {JCAP}\ }\textbf {\bibinfo {volume} {11}},\ \bibinfo
  {pages} {013} (\bibinfo {year} {2019})},\ \Eprint
  {http://arxiv.org/abs/1909.04140} {arXiv:1909.04140 [astro-ph.GA]}
  \BibitemShut {NoStop}%
\bibitem [{\citenamefont {Herrera}\ and\ \citenamefont
  {Ibarra}(2021)}]{Herrera:2021puj}%
  \BibitemOpen
  \bibfield  {author} {\bibinfo {author} {\bibfnamefont {G.}~\bibnamefont
  {Herrera}}\ and\ \bibinfo {author} {\bibfnamefont {A.}~\bibnamefont
  {Ibarra}},\ }\href {\doibase 10.1016/j.physletb.2021.136551} {\bibfield
  {journal} {\bibinfo  {journal} {Phys. Lett. B}\ }\textbf {\bibinfo {volume}
  {820}},\ \bibinfo {pages} {136551} (\bibinfo {year} {2021})},\ \Eprint
  {http://arxiv.org/abs/2104.04445} {arXiv:2104.04445 [hep-ph]} \BibitemShut
  {NoStop}%
\bibitem [{\citenamefont {Sato}\ and\ \citenamefont
  {Hughes}(2017)}]{Sato:2016slx}%
  \BibitemOpen
  \bibfield  {author} {\bibinfo {author} {\bibfnamefont {T.}~\bibnamefont
  {Sato}}\ and\ \bibinfo {author} {\bibfnamefont {J.~P.}\ \bibnamefont
  {Hughes}},\ }\href {\doibase 10.3847/1538-4357/aa6f60} {\bibfield  {journal}
  {\bibinfo  {journal} {Astrophys. J.}\ }\textbf {\bibinfo {volume} {840}},\
  \bibinfo {pages} {112} (\bibinfo {year} {2017})},\ \Eprint
  {http://arxiv.org/abs/1605.09059} {arXiv:1605.09059 [astro-ph.HE]}
  \BibitemShut {NoStop}%
\bibitem [{\citenamefont {Fitzpatrick}\ \emph {et~al.}(2013)\citenamefont
  {Fitzpatrick}, \citenamefont {Haxton}, \citenamefont {Katz}, \citenamefont
  {Lubbers},\ and\ \citenamefont {Xu}}]{Fitzpatrick:2012ix}%
  \BibitemOpen
  \bibfield  {author} {\bibinfo {author} {\bibfnamefont {A.~L.}\ \bibnamefont
  {Fitzpatrick}}, \bibinfo {author} {\bibfnamefont {W.}~\bibnamefont {Haxton}},
  \bibinfo {author} {\bibfnamefont {E.}~\bibnamefont {Katz}}, \bibinfo {author}
  {\bibfnamefont {N.}~\bibnamefont {Lubbers}}, \ and\ \bibinfo {author}
  {\bibfnamefont {Y.}~\bibnamefont {Xu}},\ }\href {\doibase
  10.1088/1475-7516/2013/02/004} {\bibfield  {journal} {\bibinfo  {journal}
  {JCAP}\ }\textbf {\bibinfo {volume} {02}},\ \bibinfo {pages} {004} (\bibinfo
  {year} {2013})},\ \Eprint {http://arxiv.org/abs/1203.3542} {arXiv:1203.3542
  [hep-ph]} \BibitemShut {NoStop}%
\bibitem [{\citenamefont {Anand}\ \emph {et~al.}(2014)\citenamefont {Anand},
  \citenamefont {Fitzpatrick},\ and\ \citenamefont {Haxton}}]{Anand:2013yka}%
  \BibitemOpen
  \bibfield  {author} {\bibinfo {author} {\bibfnamefont {N.}~\bibnamefont
  {Anand}}, \bibinfo {author} {\bibfnamefont {A.~L.}\ \bibnamefont
  {Fitzpatrick}}, \ and\ \bibinfo {author} {\bibfnamefont {W.~C.}\ \bibnamefont
  {Haxton}},\ }\href {\doibase 10.1103/PhysRevC.89.065501} {\bibfield
  {journal} {\bibinfo  {journal} {Phys. Rev. C}\ }\textbf {\bibinfo {volume}
  {89}},\ \bibinfo {pages} {065501} (\bibinfo {year} {2014})},\ \Eprint
  {http://arxiv.org/abs/1308.6288} {arXiv:1308.6288 [hep-ph]} \BibitemShut
  {NoStop}%
\bibitem [{\citenamefont {Jungman}\ \emph {et~al.}(1996)\citenamefont
  {Jungman}, \citenamefont {Kamionkowski},\ and\ \citenamefont
  {Griest}}]{Jungman:1995df}%
  \BibitemOpen
  \bibfield  {author} {\bibinfo {author} {\bibfnamefont {G.}~\bibnamefont
  {Jungman}}, \bibinfo {author} {\bibfnamefont {M.}~\bibnamefont
  {Kamionkowski}}, \ and\ \bibinfo {author} {\bibfnamefont {K.}~\bibnamefont
  {Griest}},\ }\href {\doibase 10.1016/0370-1573(95)00058-5} {\bibfield
  {journal} {\bibinfo  {journal} {Phys. Rept.}\ }\textbf {\bibinfo {volume}
  {267}},\ \bibinfo {pages} {195} (\bibinfo {year} {1996})},\ \Eprint
  {http://arxiv.org/abs/hep-ph/9506380} {arXiv:hep-ph/9506380} \BibitemShut
  {NoStop}%
\bibitem [{\citenamefont {Chang}\ \emph {et~al.}(2010)\citenamefont {Chang},
  \citenamefont {Pierce},\ and\ \citenamefont {Weiner}}]{Chang:2009yt}%
  \BibitemOpen
  \bibfield  {author} {\bibinfo {author} {\bibfnamefont {S.}~\bibnamefont
  {Chang}}, \bibinfo {author} {\bibfnamefont {A.}~\bibnamefont {Pierce}}, \
  and\ \bibinfo {author} {\bibfnamefont {N.}~\bibnamefont {Weiner}},\ }\href
  {\doibase 10.1088/1475-7516/2010/01/006} {\bibfield  {journal} {\bibinfo
  {journal} {JCAP}\ }\textbf {\bibinfo {volume} {01}},\ \bibinfo {pages} {006}
  (\bibinfo {year} {2010})},\ \Eprint {http://arxiv.org/abs/0908.3192}
  {arXiv:0908.3192 [hep-ph]} \BibitemShut {NoStop}%
\bibitem [{\citenamefont {Fan}\ \emph {et~al.}(2010)\citenamefont {Fan},
  \citenamefont {Reece},\ and\ \citenamefont {Wang}}]{Fan:2010gt}%
  \BibitemOpen
  \bibfield  {author} {\bibinfo {author} {\bibfnamefont {J.}~\bibnamefont
  {Fan}}, \bibinfo {author} {\bibfnamefont {M.}~\bibnamefont {Reece}}, \ and\
  \bibinfo {author} {\bibfnamefont {L.-T.}\ \bibnamefont {Wang}},\ }\href
  {\doibase 10.1088/1475-7516/2010/11/042} {\bibfield  {journal} {\bibinfo
  {journal} {JCAP}\ }\textbf {\bibinfo {volume} {11}},\ \bibinfo {pages} {042}
  (\bibinfo {year} {2010})},\ \Eprint {http://arxiv.org/abs/1008.1591}
  {arXiv:1008.1591 [hep-ph]} \BibitemShut {NoStop}%
\bibitem [{\citenamefont {Kumar}\ and\ \citenamefont
  {Marfatia}(2013)}]{Kumar:2013iva}%
  \BibitemOpen
  \bibfield  {author} {\bibinfo {author} {\bibfnamefont {J.}~\bibnamefont
  {Kumar}}\ and\ \bibinfo {author} {\bibfnamefont {D.}~\bibnamefont
  {Marfatia}},\ }\href {\doibase 10.1103/PhysRevD.88.014035} {\bibfield
  {journal} {\bibinfo  {journal} {Phys. Rev. D}\ }\textbf {\bibinfo {volume}
  {88}},\ \bibinfo {pages} {014035} (\bibinfo {year} {2013})},\ \Eprint
  {http://arxiv.org/abs/1305.1611} {arXiv:1305.1611 [hep-ph]} \BibitemShut
  {NoStop}%
\bibitem [{\citenamefont {Catena}\ \emph {et~al.}(2019)\citenamefont {Catena},
  \citenamefont {Fridell},\ and\ \citenamefont {Krauss}}]{Catena:2019hzw}%
  \BibitemOpen
  \bibfield  {author} {\bibinfo {author} {\bibfnamefont {R.}~\bibnamefont
  {Catena}}, \bibinfo {author} {\bibfnamefont {K.}~\bibnamefont {Fridell}}, \
  and\ \bibinfo {author} {\bibfnamefont {M.~B.}\ \bibnamefont {Krauss}},\
  }\href {\doibase 10.1007/JHEP08(2019)030} {\bibfield  {journal} {\bibinfo
  {journal} {JHEP}\ }\textbf {\bibinfo {volume} {08}},\ \bibinfo {pages} {030}
  (\bibinfo {year} {2019})},\ \Eprint {http://arxiv.org/abs/1907.02910}
  {arXiv:1907.02910 [hep-ph]} \BibitemShut {NoStop}%
\bibitem [{\citenamefont {Gondolo}\ \emph {et~al.}(2021)\citenamefont
  {Gondolo}, \citenamefont {Kang}, \citenamefont {Scopel},\ and\ \citenamefont
  {Tomar}}]{Gondolo:2020wge}%
  \BibitemOpen
  \bibfield  {author} {\bibinfo {author} {\bibfnamefont {P.}~\bibnamefont
  {Gondolo}}, \bibinfo {author} {\bibfnamefont {S.}~\bibnamefont {Kang}},
  \bibinfo {author} {\bibfnamefont {S.}~\bibnamefont {Scopel}}, \ and\ \bibinfo
  {author} {\bibfnamefont {G.}~\bibnamefont {Tomar}},\ }\href {\doibase
  10.1103/PhysRevD.104.063017} {\bibfield  {journal} {\bibinfo  {journal}
  {Phys. Rev. D}\ }\textbf {\bibinfo {volume} {104}},\ \bibinfo {pages}
  {063017} (\bibinfo {year} {2021})},\ \Eprint
  {http://arxiv.org/abs/2008.05120} {arXiv:2008.05120 [hep-ph]} \BibitemShut
  {NoStop}%
\bibitem [{\citenamefont {Catena}\ and\ \citenamefont
  {Schwabe}(2015)}]{Catena:2015uha}%
  \BibitemOpen
  \bibfield  {author} {\bibinfo {author} {\bibfnamefont {R.}~\bibnamefont
  {Catena}}\ and\ \bibinfo {author} {\bibfnamefont {B.}~\bibnamefont
  {Schwabe}},\ }\href {\doibase 10.1088/1475-7516/2015/04/042} {\bibfield
  {journal} {\bibinfo  {journal} {JCAP}\ }\textbf {\bibinfo {volume} {04}},\
  \bibinfo {pages} {042} (\bibinfo {year} {2015})},\ \Eprint
  {http://arxiv.org/abs/1501.03729} {arXiv:1501.03729 [hep-ph]} \BibitemShut
  {NoStop}%
\bibitem [{\citenamefont {{Sedov}}(1946)}]{1946JApMM..10..241S}%
  \BibitemOpen
  \bibfield  {author} {\bibinfo {author} {\bibfnamefont {L.~I.}\ \bibnamefont
  {{Sedov}}},\ }\href@noop {} {\bibfield  {journal} {\bibinfo  {journal}
  {Journal of Applied Mathematics and Mechanics}\ }\textbf {\bibinfo {volume}
  {10}},\ \bibinfo {pages} {241} (\bibinfo {year} {1946})}\BibitemShut
  {NoStop}%
\bibitem [{\citenamefont {{Taylor}}(1950{\natexlab{a}})}]{1950RSPSA.201..159T}%
  \BibitemOpen
  \bibfield  {author} {\bibinfo {author} {\bibfnamefont {G.}~\bibnamefont
  {{Taylor}}},\ }\href {\doibase 10.1098/rspa.1950.0049} {\bibfield  {journal}
  {\bibinfo  {journal} {Proceedings of the Royal Society of London Series A}\
  }\textbf {\bibinfo {volume} {201}},\ \bibinfo {pages} {159} (\bibinfo {year}
  {1950}{\natexlab{a}})}\BibitemShut {NoStop}%
\bibitem [{\citenamefont {{Taylor}}(1950{\natexlab{b}})}]{1950RSPSA.201..175T}%
  \BibitemOpen
  \bibfield  {author} {\bibinfo {author} {\bibfnamefont {G.}~\bibnamefont
  {{Taylor}}},\ }\href {\doibase 10.1098/rspa.1950.0050} {\bibfield  {journal}
  {\bibinfo  {journal} {Proceedings of the Royal Society of London Series A}\
  }\textbf {\bibinfo {volume} {201}},\ \bibinfo {pages} {175} (\bibinfo {year}
  {1950}{\natexlab{b}})}\BibitemShut {NoStop}%
\bibitem [{\citenamefont {{Sedov}}(1959)}]{1959sdmm.book.....S}%
  \BibitemOpen
  \bibfield  {author} {\bibinfo {author} {\bibfnamefont {L.~I.}\ \bibnamefont
  {{Sedov}}},\ }\href@noop {} {\emph {\bibinfo {title} {{Similarity and
  Dimensional Methods in Mechanics}}}}\ (\bibinfo {year} {1959})\BibitemShut
  {NoStop}%
\bibitem [{\citenamefont {Cardillo}\ \emph {et~al.}(2015)\citenamefont
  {Cardillo}, \citenamefont {Amato},\ and\ \citenamefont
  {Blasi}}]{Cardillo:2015zda}%
  \BibitemOpen
  \bibfield  {author} {\bibinfo {author} {\bibfnamefont {M.}~\bibnamefont
  {Cardillo}}, \bibinfo {author} {\bibfnamefont {E.}~\bibnamefont {Amato}}, \
  and\ \bibinfo {author} {\bibfnamefont {P.}~\bibnamefont {Blasi}},\ }\href
  {\doibase 10.1016/j.astropartphys.2015.03.002} {\bibfield  {journal}
  {\bibinfo  {journal} {Astropart. Phys.}\ }\textbf {\bibinfo {volume} {69}},\
  \bibinfo {pages} {1} (\bibinfo {year} {2015})},\ \Eprint
  {http://arxiv.org/abs/1503.03001} {arXiv:1503.03001 [astro-ph.HE]}
  \BibitemShut {NoStop}%
\bibitem [{\citenamefont {Cristofari}\ \emph {et~al.}(2021)\citenamefont
  {Cristofari}, \citenamefont {Blasi},\ and\ \citenamefont
  {Caprioli}}]{Cristofari:2021hbc}%
  \BibitemOpen
  \bibfield  {author} {\bibinfo {author} {\bibfnamefont {P.}~\bibnamefont
  {Cristofari}}, \bibinfo {author} {\bibfnamefont {P.}~\bibnamefont {Blasi}}, \
  and\ \bibinfo {author} {\bibfnamefont {D.}~\bibnamefont {Caprioli}},\ }\href
  {\doibase 10.1051/0004-6361/202140448} {\bibfield  {journal} {\bibinfo
  {journal} {Astron. Astrophys.}\ }\textbf {\bibinfo {volume} {650}},\ \bibinfo
  {pages} {A62} (\bibinfo {year} {2021})},\ \Eprint
  {http://arxiv.org/abs/2103.02375} {arXiv:2103.02375 [astro-ph.HE]}
  \BibitemShut {NoStop}%
\bibitem [{\citenamefont {Rauscher}\ \emph {et~al.}(2002)\citenamefont
  {Rauscher}, \citenamefont {Heger}, \citenamefont {Hoffman},\ and\
  \citenamefont {Woosley}}]{Rauscher:2001dw}%
  \BibitemOpen
  \bibfield  {author} {\bibinfo {author} {\bibfnamefont {T.}~\bibnamefont
  {Rauscher}}, \bibinfo {author} {\bibfnamefont {A.}~\bibnamefont {Heger}},
  \bibinfo {author} {\bibfnamefont {R.~D.}\ \bibnamefont {Hoffman}}, \ and\
  \bibinfo {author} {\bibfnamefont {S.~E.}\ \bibnamefont {Woosley}},\ }\href
  {\doibase 10.1086/341728} {\bibfield  {journal} {\bibinfo  {journal}
  {Astrophys. J.}\ }\textbf {\bibinfo {volume} {576}},\ \bibinfo {pages} {323}
  (\bibinfo {year} {2002})},\ \Eprint {http://arxiv.org/abs/astro-ph/0112478}
  {arXiv:astro-ph/0112478} \BibitemShut {NoStop}%
\bibitem [{\citenamefont {{Plucinsky}}\ \emph {et~al.}(1996)\citenamefont
  {{Plucinsky}}, \citenamefont {{Snowden}}, \citenamefont {{Aschenbach}},
  \citenamefont {{Egger}}, \citenamefont {{Edgar}},\ and\ \citenamefont
  {{McCammon}}}]{1996ApJ...463..224P}%
  \BibitemOpen
  \bibfield  {author} {\bibinfo {author} {\bibfnamefont {P.~P.}\ \bibnamefont
  {{Plucinsky}}}, \bibinfo {author} {\bibfnamefont {S.~L.}\ \bibnamefont
  {{Snowden}}}, \bibinfo {author} {\bibfnamefont {B.}~\bibnamefont
  {{Aschenbach}}}, \bibinfo {author} {\bibfnamefont {R.}~\bibnamefont
  {{Egger}}}, \bibinfo {author} {\bibfnamefont {R.~J.}\ \bibnamefont
  {{Edgar}}}, \ and\ \bibinfo {author} {\bibfnamefont {D.}~\bibnamefont
  {{McCammon}}},\ }\href {\doibase 10.1086/177236} {\bibfield  {journal}
  {\bibinfo  {journal} {\apj}\ }\textbf {\bibinfo {volume} {463}},\ \bibinfo
  {pages} {224} (\bibinfo {year} {1996})}\BibitemShut {NoStop}%
\bibitem [{\citenamefont {Thorsett}\ \emph {et~al.}(2003)\citenamefont
  {Thorsett}, \citenamefont {Benjamin}, \citenamefont {Brisken}, \citenamefont
  {Golden},\ and\ \citenamefont {Goss}}]{Thorsett:2003xy}%
  \BibitemOpen
  \bibfield  {author} {\bibinfo {author} {\bibfnamefont {S.~E.}\ \bibnamefont
  {Thorsett}}, \bibinfo {author} {\bibfnamefont {R.~A.}\ \bibnamefont
  {Benjamin}}, \bibinfo {author} {\bibfnamefont {W.~F.}\ \bibnamefont
  {Brisken}}, \bibinfo {author} {\bibfnamefont {A.}~\bibnamefont {Golden}}, \
  and\ \bibinfo {author} {\bibfnamefont {W.~M.}\ \bibnamefont {Goss}},\ }\href
  {\doibase 10.1086/377682} {\bibfield  {journal} {\bibinfo  {journal}
  {Astrophys. J. Lett.}\ }\textbf {\bibinfo {volume} {592}},\ \bibinfo {pages}
  {L71} (\bibinfo {year} {2003})},\ \Eprint
  {http://arxiv.org/abs/astro-ph/0306462} {arXiv:astro-ph/0306462} \BibitemShut
  {NoStop}%
\bibitem [{\citenamefont {{Knies}}\ \emph {et~al.}(2018)\citenamefont
  {{Knies}}, \citenamefont {{Sasaki}},\ and\ \citenamefont
  {{Plucinsky}}}]{2018MNRAS.477.4414K}%
  \BibitemOpen
  \bibfield  {author} {\bibinfo {author} {\bibfnamefont {J.~R.}\ \bibnamefont
  {{Knies}}}, \bibinfo {author} {\bibfnamefont {M.}~\bibnamefont {{Sasaki}}}, \
  and\ \bibinfo {author} {\bibfnamefont {P.~P.}\ \bibnamefont {{Plucinsky}}},\
  }\href {\doibase 10.1093/mnras/sty915} {\bibfield  {journal} {\bibinfo
  {journal} {\mnras}\ }\textbf {\bibinfo {volume} {477}},\ \bibinfo {pages}
  {4414} (\bibinfo {year} {2018})}\BibitemShut {NoStop}%
\bibitem [{\citenamefont {{Chevalier}}(1982)}]{1982ApJ...258..790C}%
  \BibitemOpen
  \bibfield  {author} {\bibinfo {author} {\bibfnamefont {R.~A.}\ \bibnamefont
  {{Chevalier}}},\ }\href {\doibase 10.1086/160126} {\bibfield  {journal}
  {\bibinfo  {journal} {\apj}\ }\textbf {\bibinfo {volume} {258}},\ \bibinfo
  {pages} {790} (\bibinfo {year} {1982})}\BibitemShut {NoStop}%
\bibitem [{\citenamefont {Athron}\ \emph {et~al.}(2021)\citenamefont {Athron}
  \emph {et~al.}}]{GAMBIT:2021rlp}%
  \BibitemOpen
  \bibfield  {author} {\bibinfo {author} {\bibfnamefont {P.}~\bibnamefont
  {Athron}} \emph {et~al.} (\bibinfo {collaboration} {GAMBIT}),\ }\href
  {\doibase 10.1140/epjc/s10052-021-09712-6} {\bibfield  {journal} {\bibinfo
  {journal} {Eur. Phys. J. C}\ }\textbf {\bibinfo {volume} {81}},\ \bibinfo
  {pages} {992} (\bibinfo {year} {2021})},\ \Eprint
  {http://arxiv.org/abs/2106.02056} {arXiv:2106.02056 [hep-ph]} \BibitemShut
  {NoStop}%
\bibitem [{\citenamefont {Kozar}\ \emph {et~al.}(2021)\citenamefont {Kozar},
  \citenamefont {Caddell}, \citenamefont {Fraser-Leach}, \citenamefont
  {Scott},\ and\ \citenamefont {Vincent}}]{Kozar:2021iur}%
  \BibitemOpen
  \bibfield  {author} {\bibinfo {author} {\bibfnamefont {N.~A.}\ \bibnamefont
  {Kozar}}, \bibinfo {author} {\bibfnamefont {A.}~\bibnamefont {Caddell}},
  \bibinfo {author} {\bibfnamefont {L.}~\bibnamefont {Fraser-Leach}}, \bibinfo
  {author} {\bibfnamefont {P.}~\bibnamefont {Scott}}, \ and\ \bibinfo {author}
  {\bibfnamefont {A.~C.}\ \bibnamefont {Vincent}},\ }in\ \href@noop {} {\emph
  {\bibinfo {booktitle} {{Tools for High Energy Physics and Cosmology}}}}\
  (\bibinfo {year} {2021})\ \Eprint {http://arxiv.org/abs/2105.06810}
  {arXiv:2105.06810 [hep-ph]} \BibitemShut {NoStop}%
\bibitem [{\citenamefont {Kavanagh}\ and\ \citenamefont
  {Edwards}(2018)}]{WIMpy-code}%
  \BibitemOpen
  \bibfield  {author} {\bibinfo {author} {\bibfnamefont {B.~J.}\ \bibnamefont
  {Kavanagh}}\ and\ \bibinfo {author} {\bibfnamefont {T.~D.~P.}\ \bibnamefont
  {Edwards}},\ }\href@noop {} {\enquote {\bibinfo {title}
  {\textnormal{WIMpy\_NREFT v1.1 [Computer Software]},
  \href{https://doi.org/10.5281/zenodo.1230503}{\textnormal{doi:10.5281/zenodo.1230503}}\textnormal{.
  Available at }\url{https://github.com/bradkav/WIMpy_NREFT}},}\ } (\bibinfo
  {year} {2018})\BibitemShut {NoStop}%
\bibitem [{\citenamefont {Adhikari}\ \emph {et~al.}(2020)\citenamefont
  {Adhikari} \emph {et~al.}}]{DEAP:2020iwi}%
  \BibitemOpen
  \bibfield  {author} {\bibinfo {author} {\bibfnamefont {P.}~\bibnamefont
  {Adhikari}} \emph {et~al.} (\bibinfo {collaboration} {DEAP}),\ }\href
  {\doibase 10.1103/PhysRevD.102.082001} {\bibfield  {journal} {\bibinfo
  {journal} {Phys. Rev. D}\ }\textbf {\bibinfo {volume} {102}},\ \bibinfo
  {pages} {082001} (\bibinfo {year} {2020})},\ \bibinfo {note} {[Erratum:
  Phys.Rev.D 105, 029901(E) (2022)]},\ \Eprint
  {http://arxiv.org/abs/2005.14667} {arXiv:2005.14667 [astro-ph.CO]}
  \BibitemShut {NoStop}%
\bibitem [{\citenamefont {Amole}\ \emph {et~al.}(2019)\citenamefont {Amole}
  \emph {et~al.}}]{PICO:2019vsc}%
  \BibitemOpen
  \bibfield  {author} {\bibinfo {author} {\bibfnamefont {C.}~\bibnamefont
  {Amole}} \emph {et~al.} (\bibinfo {collaboration} {PICO}),\ }\href {\doibase
  10.1103/PhysRevD.100.022001} {\bibfield  {journal} {\bibinfo  {journal}
  {Phys. Rev. D}\ }\textbf {\bibinfo {volume} {100}},\ \bibinfo {pages}
  {022001} (\bibinfo {year} {2019})},\ \Eprint
  {http://arxiv.org/abs/1902.04031} {arXiv:1902.04031 [astro-ph.CO]}
  \BibitemShut {NoStop}%
\bibitem [{\citenamefont {Agnese}\ \emph {et~al.}(2017)\citenamefont {Agnese}
  \emph {et~al.}}]{SuperCDMS:2016wui}%
  \BibitemOpen
  \bibfield  {author} {\bibinfo {author} {\bibfnamefont {R.}~\bibnamefont
  {Agnese}} \emph {et~al.} (\bibinfo {collaboration} {SuperCDMS}),\ }\href
  {\doibase 10.1103/PhysRevD.95.082002} {\bibfield  {journal} {\bibinfo
  {journal} {Phys. Rev. D}\ }\textbf {\bibinfo {volume} {95}},\ \bibinfo
  {pages} {082002} (\bibinfo {year} {2017})},\ \Eprint
  {http://arxiv.org/abs/1610.00006} {arXiv:1610.00006 [physics.ins-det]}
  \BibitemShut {NoStop}%
\bibitem [{\citenamefont {Aalbers}\ \emph {et~al.}(2016)\citenamefont {Aalbers}
  \emph {et~al.}}]{DARWIN:2016hyl}%
  \BibitemOpen
  \bibfield  {author} {\bibinfo {author} {\bibfnamefont {J.}~\bibnamefont
  {Aalbers}} \emph {et~al.} (\bibinfo {collaboration} {DARWIN}),\ }\href
  {\doibase 10.1088/1475-7516/2016/11/017} {\bibfield  {journal} {\bibinfo
  {journal} {JCAP}\ }\textbf {\bibinfo {volume} {11}},\ \bibinfo {pages} {017}
  (\bibinfo {year} {2016})},\ \Eprint {http://arxiv.org/abs/1606.07001}
  {arXiv:1606.07001 [astro-ph.IM]} \BibitemShut {NoStop}%
\bibitem [{\citenamefont {Aprile}\ \emph {et~al.}(2018)\citenamefont {Aprile}
  \emph {et~al.}}]{XENON:2018voc}%
  \BibitemOpen
  \bibfield  {author} {\bibinfo {author} {\bibfnamefont {E.}~\bibnamefont
  {Aprile}} \emph {et~al.} (\bibinfo {collaboration} {XENON}),\ }\href
  {\doibase 10.1103/PhysRevLett.121.111302} {\bibfield  {journal} {\bibinfo
  {journal} {Phys. Rev. Lett.}\ }\textbf {\bibinfo {volume} {121}},\ \bibinfo
  {pages} {111302} (\bibinfo {year} {2018})},\ \Eprint
  {http://arxiv.org/abs/1805.12562} {arXiv:1805.12562 [astro-ph.CO]}
  \BibitemShut {NoStop}%
\bibitem [{\citenamefont {Emken}\ and\ \citenamefont
  {Kouvaris}(2018)}]{Emken:2018run}%
  \BibitemOpen
  \bibfield  {author} {\bibinfo {author} {\bibfnamefont {T.}~\bibnamefont
  {Emken}}\ and\ \bibinfo {author} {\bibfnamefont {C.}~\bibnamefont
  {Kouvaris}},\ }\href {\doibase 10.1103/PhysRevD.97.115047} {\bibfield
  {journal} {\bibinfo  {journal} {Phys. Rev. D}\ }\textbf {\bibinfo {volume}
  {97}},\ \bibinfo {pages} {115047} (\bibinfo {year} {2018})},\ \Eprint
  {http://arxiv.org/abs/1802.04764} {arXiv:1802.04764 [hep-ph]} \BibitemShut
  {NoStop}%
\bibitem [{\citenamefont {Kavanagh}(2016)}]{verne}%
  \BibitemOpen
  \bibfield  {author} {\bibinfo {author} {\bibfnamefont {B.~J.}\ \bibnamefont
  {Kavanagh}},\ }\href@noop {} {}\bibinfo {howpublished} {Verne, Astrophysics
  Source Code Library, record ascl:1802.005, available at
  \url{https://github.com/bradkav/verne}} (\bibinfo {year} {2016})\BibitemShut
  {NoStop}%
\bibitem [{\citenamefont {Kavanagh}(2018)}]{Kavanagh:2017cru}%
  \BibitemOpen
  \bibfield  {author} {\bibinfo {author} {\bibfnamefont {B.~J.}\ \bibnamefont
  {Kavanagh}},\ }\href {\doibase 10.1103/PhysRevD.97.123013} {\bibfield
  {journal} {\bibinfo  {journal} {Phys. Rev. D}\ }\textbf {\bibinfo {volume}
  {97}},\ \bibinfo {pages} {123013} (\bibinfo {year} {2018})},\ \Eprint
  {http://arxiv.org/abs/1712.04901} {arXiv:1712.04901 [hep-ph]} \BibitemShut
  {NoStop}%
\bibitem [{\citenamefont {Angloher}\ \emph {et~al.}(2019)\citenamefont
  {Angloher} \emph {et~al.}}]{CRESST:2018vwt}%
  \BibitemOpen
  \bibfield  {author} {\bibinfo {author} {\bibfnamefont {G.}~\bibnamefont
  {Angloher}} \emph {et~al.} (\bibinfo {collaboration} {CRESST}),\ }\href
  {\doibase 10.1140/epjc/s10052-018-6523-4} {\bibfield  {journal} {\bibinfo
  {journal} {Eur. Phys. J. C}\ }\textbf {\bibinfo {volume} {79}},\ \bibinfo
  {pages} {43} (\bibinfo {year} {2019})},\ \Eprint
  {http://arxiv.org/abs/1809.03753} {arXiv:1809.03753 [hep-ph]} \BibitemShut
  {NoStop}%
\bibitem [{\citenamefont {Schneck}\ \emph {et~al.}(2015)\citenamefont {Schneck}
  \emph {et~al.}}]{SuperCDMS:2015lcz}%
  \BibitemOpen
  \bibfield  {author} {\bibinfo {author} {\bibfnamefont {K.}~\bibnamefont
  {Schneck}} \emph {et~al.} (\bibinfo {collaboration} {SuperCDMS}),\ }\href
  {\doibase 10.1103/PhysRevD.91.092004} {\bibfield  {journal} {\bibinfo
  {journal} {Phys. Rev. D}\ }\textbf {\bibinfo {volume} {91}},\ \bibinfo
  {pages} {092004} (\bibinfo {year} {2015})},\ \Eprint
  {http://arxiv.org/abs/1503.03379} {arXiv:1503.03379 [astro-ph.CO]}
  \BibitemShut {NoStop}%
\bibitem [{\citenamefont {Maamari}\ \emph {et~al.}(2021)\citenamefont
  {Maamari}, \citenamefont {Gluscevic}, \citenamefont {Boddy}, \citenamefont
  {Nadler},\ and\ \citenamefont {Wechsler}}]{Maamari:2020aqz}%
  \BibitemOpen
  \bibfield  {author} {\bibinfo {author} {\bibfnamefont {K.}~\bibnamefont
  {Maamari}}, \bibinfo {author} {\bibfnamefont {V.}~\bibnamefont {Gluscevic}},
  \bibinfo {author} {\bibfnamefont {K.~K.}\ \bibnamefont {Boddy}}, \bibinfo
  {author} {\bibfnamefont {E.~O.}\ \bibnamefont {Nadler}}, \ and\ \bibinfo
  {author} {\bibfnamefont {R.~H.}\ \bibnamefont {Wechsler}},\ }\href {\doibase
  10.3847/2041-8213/abd807} {\bibfield  {journal} {\bibinfo  {journal}
  {Astrophys. J. Lett.}\ }\textbf {\bibinfo {volume} {907}},\ \bibinfo {pages}
  {L46} (\bibinfo {year} {2021})},\ \Eprint {http://arxiv.org/abs/2010.02936}
  {arXiv:2010.02936 [astro-ph.CO]} \BibitemShut {NoStop}%
\bibitem [{\citenamefont {Boddy}\ and\ \citenamefont
  {Gluscevic}(2018)}]{Boddy:2018kfv}%
  \BibitemOpen
  \bibfield  {author} {\bibinfo {author} {\bibfnamefont {K.~K.}\ \bibnamefont
  {Boddy}}\ and\ \bibinfo {author} {\bibfnamefont {V.}~\bibnamefont
  {Gluscevic}},\ }\href {\doibase 10.1103/PhysRevD.98.083510} {\bibfield
  {journal} {\bibinfo  {journal} {Phys. Rev. D}\ }\textbf {\bibinfo {volume}
  {98}},\ \bibinfo {pages} {083510} (\bibinfo {year} {2018})},\ \Eprint
  {http://arxiv.org/abs/1801.08609} {arXiv:1801.08609 [astro-ph.CO]}
  \BibitemShut {NoStop}%
\bibitem [{\citenamefont {Alvey}\ \emph {et~al.}(2022)\citenamefont {Alvey},
  \citenamefont {Bringmann},\ and\ \citenamefont {Kolesova}}]{Alvey:2022pad}%
  \BibitemOpen
  \bibfield  {author} {\bibinfo {author} {\bibfnamefont {J.}~\bibnamefont
  {Alvey}}, \bibinfo {author} {\bibfnamefont {T.}~\bibnamefont {Bringmann}}, \
  and\ \bibinfo {author} {\bibfnamefont {H.}~\bibnamefont {Kolesova}},\
  }\href@noop {} {\  (\bibinfo {year} {2022})},\ \Eprint
  {http://arxiv.org/abs/2209.03360} {arXiv:2209.03360 [hep-ph]} \BibitemShut
  {NoStop}%
\bibitem [{\citenamefont {Yellin}(2002)}]{Yellin:2002xd}%
  \BibitemOpen
  \bibfield  {author} {\bibinfo {author} {\bibfnamefont {S.}~\bibnamefont
  {Yellin}},\ }\href {\doibase 10.1103/PhysRevD.66.032005} {\bibfield
  {journal} {\bibinfo  {journal} {Phys. Rev. D}\ }\textbf {\bibinfo {volume}
  {66}},\ \bibinfo {pages} {032005} (\bibinfo {year} {2002})},\ \Eprint
  {http://arxiv.org/abs/physics/0203002} {arXiv:physics/0203002} \BibitemShut
  {NoStop}%
\bibitem [{\citenamefont {Wang}\ \emph {et~al.}(2021)\citenamefont {Wang} \emph
  {et~al.}}]{CDEX:2020tkb}%
  \BibitemOpen
  \bibfield  {author} {\bibinfo {author} {\bibfnamefont {Y.}~\bibnamefont
  {Wang}} \emph {et~al.} (\bibinfo {collaboration} {CDEX}),\ }\href {\doibase
  10.1007/s11433-020-1666-8} {\bibfield  {journal} {\bibinfo  {journal} {Sci.
  China Phys. Mech. Astron.}\ }\textbf {\bibinfo {volume} {64}},\ \bibinfo
  {pages} {281011} (\bibinfo {year} {2021})},\ \Eprint
  {http://arxiv.org/abs/2007.15555} {arXiv:2007.15555 [hep-ex]} \BibitemShut
  {NoStop}%
\bibitem [{\citenamefont {Agnes}\ \emph {et~al.}(2020)\citenamefont {Agnes}
  \emph {et~al.}}]{DarkSide-50:2020swd}%
  \BibitemOpen
  \bibfield  {author} {\bibinfo {author} {\bibfnamefont {P.}~\bibnamefont
  {Agnes}} \emph {et~al.} (\bibinfo {collaboration} {DarkSide-50}),\ }\href
  {\doibase 10.1103/PhysRevD.101.062002} {\bibfield  {journal} {\bibinfo
  {journal} {Phys. Rev. D}\ }\textbf {\bibinfo {volume} {101}},\ \bibinfo
  {pages} {062002} (\bibinfo {year} {2020})},\ \Eprint
  {http://arxiv.org/abs/2002.07794} {arXiv:2002.07794 [hep-ex]} \BibitemShut
  {NoStop}%
\bibitem [{\citenamefont {Akerib}\ \emph
  {et~al.}(2021{\natexlab{a}})\citenamefont {Akerib} \emph
  {et~al.}}]{LUX:2020oan}%
  \BibitemOpen
  \bibfield  {author} {\bibinfo {author} {\bibfnamefont {D.~S.}\ \bibnamefont
  {Akerib}} \emph {et~al.} (\bibinfo {collaboration} {LUX}),\ }\href {\doibase
  10.1103/PhysRevD.103.122005} {\bibfield  {journal} {\bibinfo  {journal}
  {Phys. Rev. D}\ }\textbf {\bibinfo {volume} {103}},\ \bibinfo {pages}
  {122005} (\bibinfo {year} {2021}{\natexlab{a}})},\ \Eprint
  {http://arxiv.org/abs/2003.11141} {arXiv:2003.11141 [astro-ph.CO]}
  \BibitemShut {NoStop}%
\bibitem [{\citenamefont {Akerib}\ \emph
  {et~al.}(2021{\natexlab{b}})\citenamefont {Akerib} \emph
  {et~al.}}]{LUX:2021ksq}%
  \BibitemOpen
  \bibfield  {author} {\bibinfo {author} {\bibfnamefont {D.~S.}\ \bibnamefont
  {Akerib}} \emph {et~al.} (\bibinfo {collaboration} {LUX}),\ }\href {\doibase
  10.1103/PhysRevD.104.062005} {\bibfield  {journal} {\bibinfo  {journal}
  {Phys. Rev. D}\ }\textbf {\bibinfo {volume} {104}},\ \bibinfo {pages}
  {062005} (\bibinfo {year} {2021}{\natexlab{b}})},\ \Eprint
  {http://arxiv.org/abs/2102.06998} {arXiv:2102.06998 [astro-ph.CO]}
  \BibitemShut {NoStop}%
\bibitem [{\citenamefont {Xia}\ \emph {et~al.}(2019)\citenamefont {Xia} \emph
  {et~al.}}]{PandaX-II:2018woa}%
  \BibitemOpen
  \bibfield  {author} {\bibinfo {author} {\bibfnamefont {J.}~\bibnamefont
  {Xia}} \emph {et~al.} (\bibinfo {collaboration} {PandaX-II}),\ }\href
  {\doibase 10.1016/j.physletb.2019.02.043} {\bibfield  {journal} {\bibinfo
  {journal} {Phys. Lett. B}\ }\textbf {\bibinfo {volume} {792}},\ \bibinfo
  {pages} {193} (\bibinfo {year} {2019})},\ \Eprint
  {http://arxiv.org/abs/1807.01936} {arXiv:1807.01936 [hep-ex]} \BibitemShut
  {NoStop}%
\bibitem [{\citenamefont {Albakry}\ \emph
  {et~al.}(2022{\natexlab{b}})\citenamefont {Albakry} \emph
  {et~al.}}]{SuperCDMS:2022crd}%
  \BibitemOpen
  \bibfield  {author} {\bibinfo {author} {\bibfnamefont {M.~F.}\ \bibnamefont
  {Albakry}} \emph {et~al.} (\bibinfo {collaboration} {SuperCDMS}),\
  }\href@noop {} {\  (\bibinfo {year} {2022}{\natexlab{b}})},\ \Eprint
  {http://arxiv.org/abs/2205.11683} {arXiv:2205.11683 [astro-ph.CO]}
  \BibitemShut {NoStop}%
\bibitem [{\citenamefont {Aprile}\ \emph {et~al.}(2017)\citenamefont {Aprile}
  \emph {et~al.}}]{XENON:2017fdd}%
  \BibitemOpen
  \bibfield  {author} {\bibinfo {author} {\bibfnamefont {E.}~\bibnamefont
  {Aprile}} \emph {et~al.} (\bibinfo {collaboration} {XENON}),\ }\href
  {\doibase 10.1103/PhysRevD.96.042004} {\bibfield  {journal} {\bibinfo
  {journal} {Phys. Rev. D}\ }\textbf {\bibinfo {volume} {96}},\ \bibinfo
  {pages} {042004} (\bibinfo {year} {2017})},\ \Eprint
  {http://arxiv.org/abs/1705.02614} {arXiv:1705.02614 [astro-ph.CO]}
  \BibitemShut {NoStop}%
\end{thebibliography}%

\end{document}